\documentclass[useAMS]{mn2e}
\usepackage[tight]{subfigure}
\usepackage{longtable}
\usepackage{amssymb}
\usepackage{lscape}
\usepackage{graphicx}

\title[]{Stellar populations in central cluster galaxies: the influence of cooling flows\thanks{based on observations made with the Southern African Large Telescope (SALT).}}
\author[Loubser et al.]{S. I. Loubser$^{1}$\thanks{E-mail:
Ilani.Loubser@nwu.ac.za (SIL)}\\
$^{1}$Centre for Space Research, North-West University, Potchefstroom 2520, South Africa\\}
\begin{document}


\pagerange{\pageref{firstpage}--\pageref{lastpage}} \pubyear{2012}

\maketitle

\label{firstpage}

\begin{abstract}
We present detailed, high spatial and spectral resolution, long-slit observations of four central cluster galaxies (Abell 0085, 0133, 0644 and Ophiuchus) recently obtained on the Southern African Large Telescope (SALT). Our sample consists of central cluster galaxies (CCGs) with previously-observed H$\alpha$ filaments, and have existing data from the X-ray to radio wavelength regimes available. Here, we present the detailed optical data over a broad wavelength range to probe the spatially-resolved kinematics and stellar populations of the stars. We use the Pegase.HR model with the ELODIE v3.1 stellar library to determine the star formation histories of the galaxies using full spectrum fitting. We perform single stellar population (SSP) as well as composite stellar population (CSP) fits to account for more complex star formation histories. Monte-Carlo simulations and $\chi^{2}$-maps are used to check the reliability of the solutions. This, combined with the other multiwavelength data, will form a complete view of the different phases (hot and cold gas and stars) and how they interact in the processes of star formation and feedback detected in central galaxies in cooling flow clusters, as well as the influence of the host cluster. We find small, young stellar components in at least three of the four galaxies, even though two of the three host clusters have zero spectrally-derived mass deposition rates from X-ray observations.
\end{abstract}


\section{Introduction}

Cooling flows in galaxy clusters, once thought to be on the order of 100 -- 1000 M$_{\odot}$ yr$^{-1}$, are now understood to be depositing cool gas on the order of 1 -- 10 M$_{\odot}$ yr$^{-1}$ onto the central cluster galaxy (hereafter abbreviated to CCG; Voigt $\&$ Fabian 2004). It is now generally accepted that some form of feedback (AGN, conduction etc.) prevents the intracluster medium from cooling, allowing only small amounts of cool gas to accrete onto the CCG. Thus, CCGs lie at the interface where it is crucial to understand the role of feedback and accretion in star formation. Within these cooling-flow CCGs, cool molecular clouds, warm ionized hydrogen, and the cooling intracluster medium are related. A complete view of the star formation process incorporates the stars with the gas and an understanding of the processes by which these phases interact, and therefore, requires information from several wavelength regimes.

Various previous studies have reported several examples of ongoing star formation in CCGs, in particular those hosted by cooling-flow clusters (Cardiel, Gorgas $\&$ Arag\'{o}n-Salamanca 1998; Crawford et al.\ 1999; McNamara et al.\ 2006; Edwards et al.\ 2007; O'Dea et al.\ 2008; Bildfell et al.\ 2008; Pipino et al.\ 2009; Loubser et al.\ 2009; O'Dea et al.\ 2010; Liu, Mao $\&$ Meng 2012). Although active star formation in these central galaxies is compelling, the young populations only contribute a very small mass fraction (Pipino et al.\ 2009), and it has been shown that star formation in these CCGs is correlated with the cooling time of the gas (Rafferty, McNamara $\&$ Nulsen 2008; Liu et al.\ 2012). Recent studies have even claimed that star forming central galaxies are exclusively hosted by cooling flow clusters (Hoffer et al.\ 2012, and references therein), although not all cooling-flow clusters contain star forming galaxies. The empirical boundary between clusters that host active central galaxies and clusters that never host them is $K_{0}=30$ keV cm$^{2}$, an entropy corresponding to an intracluster medium cooling time of $\sim 1$ Gyr (Voit et al.\ 2008). The origin of the gas fuelling this star formation is not yet fully understood. Probable explanations include processes involving the cooling flows, but cold gas deposited during a merging event can not be conclusively eliminated (Bildfell et al.\ 2008).

With the advent of more accurate methods to fit complex star formation histories (SFH) from high-quality optical spectroscopy, we are now in a position to directly compare the light/mass fraction of any possible present young stellar population component to the cooling rate of the host cluster.

We proposed and obtained long-slit observations of CCGs with confirmed H$\alpha$-filaments. We selected galaxies with near-IR (2MASS), ultraviolet (GALEX), X-ray data (Chandra), and VLA 1.4 GHz fluxes, already available (McDonald et al.\ 2010). Detailed properties of the host clusters, which are reported to influence the activity in the central galaxy (as described above), such as central cooling times and the offset between the cluster X-ray peak and the central galaxy, have been derived in previous studies and are available in the literature. 

This data is complimentary to the emission-line, long-slit spectroscopy (on KECK and Magellan) along the H$\alpha$ filaments of nine CCGs (not including the four studied here) by McDonald et al.\ (2012). To get maximum possible information, we observed line ratios over a very long wavelength range (around H$\alpha$ and H$\beta$) with the slit aligned on, or as close as possible, to the major axis of the galaxy. We can now place the derived information from the stellar population analysis of the optical spectra in context with multiwavelength data over the full spectrum, to explain the diverse nature of these galaxies. 

We introduce the sample and detail of the data reductions in Sections \ref{Sample} and \ref{reduction}. We then derive the kinematics as well as the stellar populations in Section \ref{extinction}. We proceed to discuss the four individual cases in Section \ref{figures_individual}. We summarise the findings of this paper in Section \ref{summary}. We have used the following set of cosmological parameters: $\Omega_{m}$ = 0.3, $\Omega_{\Lambda}$ = 0.7, H$_{0}$ = 70km s$^{-1}$ Mpc$^{-1}$. 

\section{Sample}
\label{Sample}

\begin{figure*}
   \centering
 \mbox{\subfigure[ESO541-013]{\includegraphics[width=4.2cm,height=4.2cm]{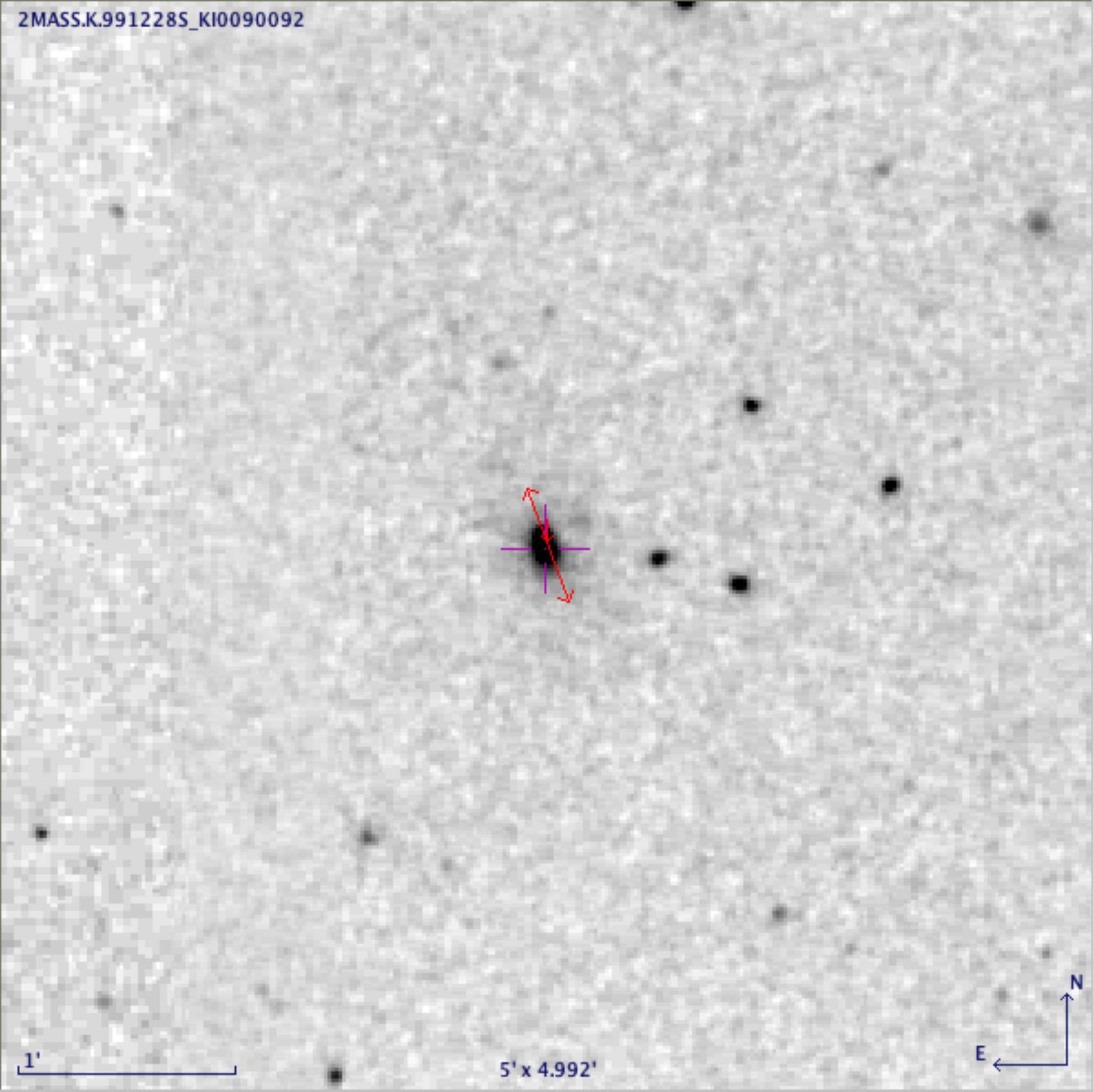}}\quad
         \subfigure[MCG02-02-086]{\includegraphics[width=4.2cm,height=4.2cm]{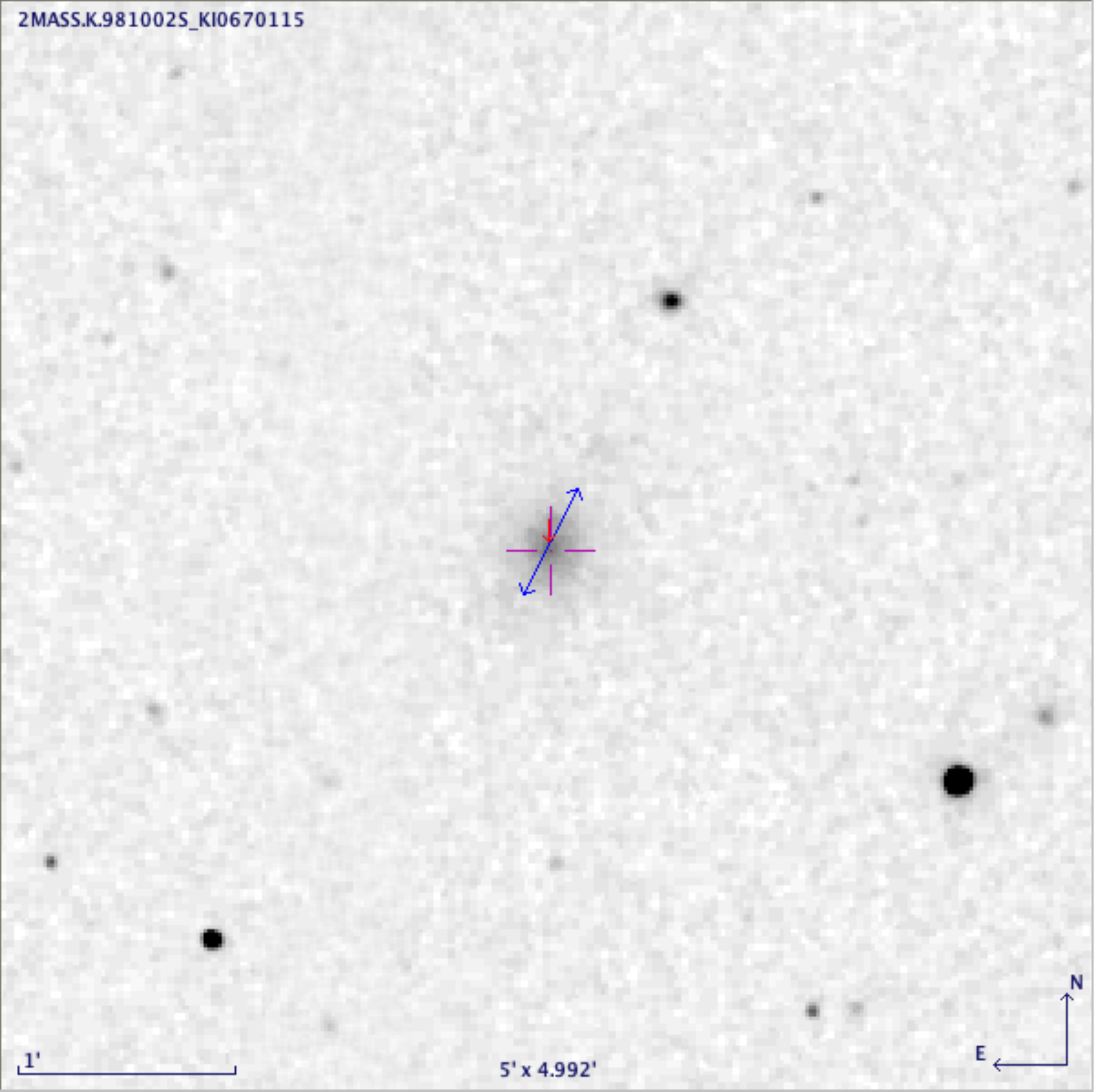}}\quad
         \subfigure[PGC023233]{\includegraphics[width=4.2cm,height=4.2cm]{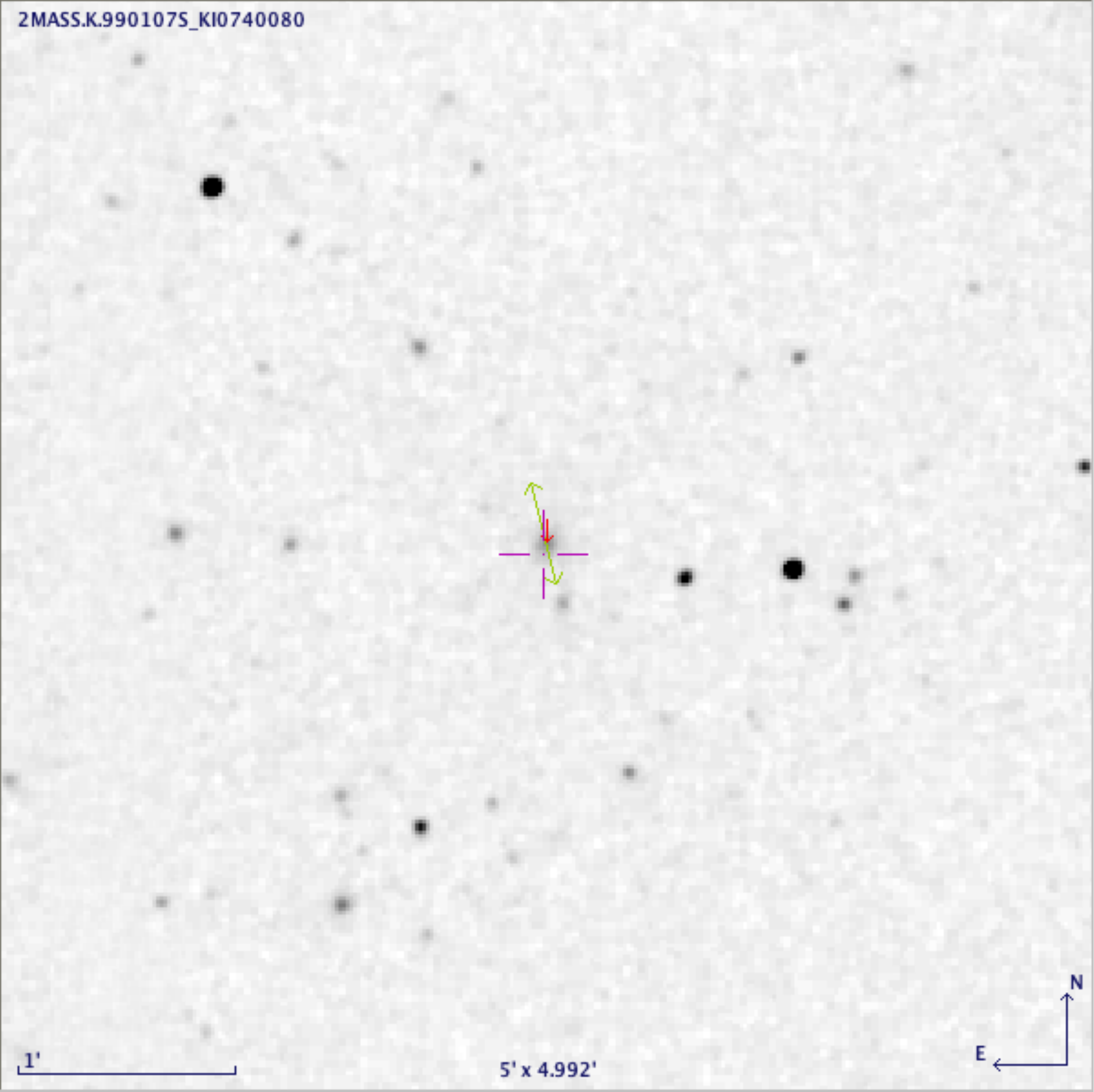}}\quad
\subfigure[2MASXJ17122774-2322108]{\includegraphics[width=4.2cm,height=4.2cm]{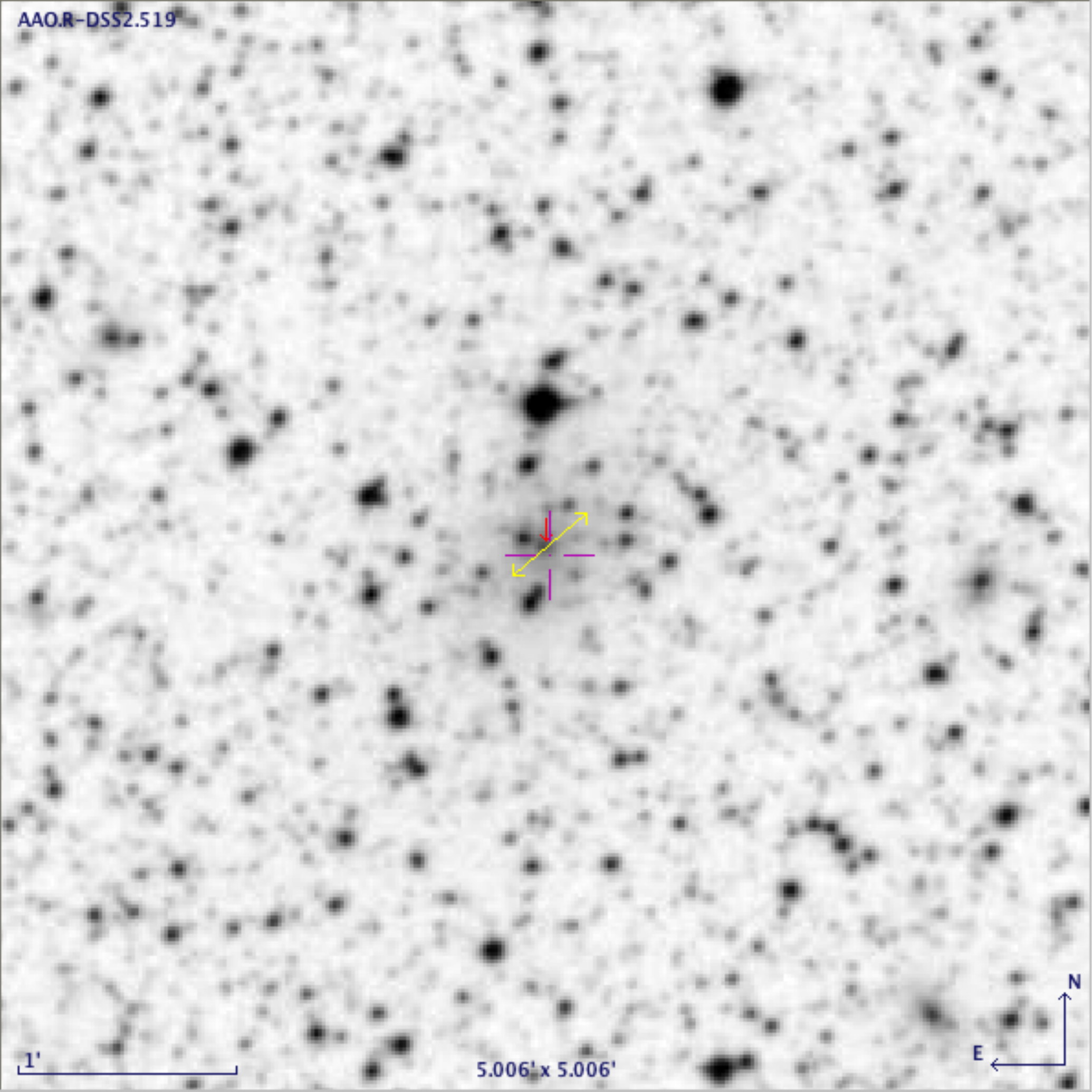}}}
\caption{Slit placements overlaid on DSS images of the four targets (E to the left). The plots show 5 $\times$ 5 arcmin field of views.}
\label{fig:Thumbnails} 
\end{figure*}

We have chosen our sample of CCGs from the H$\alpha$ imaging presented in McDonald et al. (2010), who in turn, selected their sample from White, Jones $\&$ Forman (1997). McDonald et al.\ (2010) enforced the cuts: $\delta < +35 \deg$ and $0.025 < z < 0.092$, after which they selected 23 clusters to cover the full range of properties, from very rich clusters with high cooling rates to low-density clusters with very small cooling flows. From their 23 clusters, we selected all the clusters with clearly detected H$\alpha$ in their centres (albeit filamentary, extended or nuclear emission). In addition, all of these central galaxies have optical imaging, near-IR (2MASS) and UV (Galex data) available. Thereafter, we selected all the central galaxies with detailed X-ray (Chandra) data, as well as VLA 1.4 GHz fluxes, available. This resulted in a sub-sample of 10 galaxies. We observed four of these galaxies with the GMOS IFU (as presented in Loubser $\&$ Soechting 2013), and we observed four of these galaxies with the SALT RSS (as shown in Figure \ref{fig:Thumbnails}). This does not constitute a complete sample as we merely chose the objects with the most auxiliary information available. 

\begin{table*}
\begin{footnotesize}
\begin{tabular}{l c c c c c c c} 
\hline Object & Cluster & Redshift & R$_{off}$ & T$_{X}$ & Classical cooling rates & Spectrally determined & Exposure time \\
       &  &  \multicolumn{1}{c}{$z$} & \multicolumn{1}{c}{(Mpc)} & \multicolumn{1}{c}{(keV)} & \multicolumn{1}{c}{(M$_{\odot}$/yr$^{-1}$)}& \multicolumn{1}{c}{(M$_{\odot}$/yr$^{-1}$)} &\multicolumn{1}{c}{(seconds)} \\
\hline					
ESO541-013 & Abell 0133  & 0.057 & 0.017 & 3.5 & 110 & 0.0 & 9704 \\
MCG-02-02-086 & Abell 0085  & 0.056 & 0.046 & 6.5 & 108 & 2.2 & 6353 \\
PGC023233 & Abell 0644  & 0.071 & -- & 6.5 & 136 & 1.5 & 6800 \\
2MASXJ17122774-2322108 & Ophiuchus & 0.028 & 0.019 & 8.6 & 41 & 0.0 & 14030 \\
\hline
\end{tabular}
\caption{Galaxies observed with the SALT telescope. All four galaxies show extended H$\alpha$ emission (McDonald et al.\ 2010). The cluster X-ray temperature (T$_{X}$) and classical cooling rates rates (\.{M}) are from White, Jones $\&$ Forman (1997). The spectrally determined cooling rates are from McDonald et al.\ (2010). The values for R$_{off}$ are from Edwards et al.\ (2007) for ESO541-013 and MCG02-02-086, and for 2MASXJ17122774-2322108 was calculated from information in NED (http://nedwww.ipac.caltech.edu/). This was not possible for Abell 0644, where the CCG is not in the centre of the main cluster and where the coordinates of a corresponding local X-ray maximum was not available.}
\label{table:objects}
\end{footnotesize}
\end{table*}

\begin{table*}
\centering
\begin{footnotesize}
\begin{tabular}{l c c } 
\hline Object & Extinction (mag) & Radio flux\\
       & E(B-V)$_{galactic}$ & (mJy)\\
\hline					
ESO541-013 & 0.020 & 167\\
MCG-02-02-086 & 0.034 & 57\\
PGC023233 & 0.109 & 0 \\
2MASXJ17122774-2322108 & 0.521 & 29\\
\hline
\end{tabular}
\caption{Further properties of the CCGs observed on SALT. Radio fluxes are from the NVSS (Condon et al.\ 1998). The NVSS images are shown in Figures \ref{fig:ESO_NVSS} to \ref{fig:2MAS_NVSS}.}
\label{table:objects2}
\end{footnotesize}
\end{table*}

\section{Observations and data reduction}
\label{reduction}
The data were obtained with the Robert Stobie Spectrograph (RSS; see Burgh et al.\ 2003 and Kobulnicky et al.\ 2003) on the SALT telescope between October 2011 and October 2012 (during two observing semesters and on 24 different nights during dark time under program 2011-2-RSA$\_$OTH-003 and 2012-1-RSA$\_$OTH-003; PI: Ilani Loubser). The rest wavelength of interest is 4860 -- 6731 \AA{} (redshifted to 5000 -- 7300 \AA{}). To achieve this, and to avoid losing essential lines in the ccd gaps, the pg0900 grating was used with 1.2'' slit at a carefully selected central wavelength setting. The targets and exposure times are shown in Table \ref{table:objects}. In addition to the targets, the necessary flat-field and arcs frames were also observed at regular intervals, as well as spectrophotometric standard stars for flux calibration.

The basic reductions were performed with the PySALT: SALT science pipeline\footnote{http://pysalt.salt.ac.za/} (Crawford et al.\ 2010), whilst further reductions were done in \textsc{IRAF}\footnote{\textsc{iraf} is distributed by the National Optical Astronomy Observatories, which are operated by the Association of Universities for Research in Astronomy Inc., under cooperative agreement with the National Science Foundation.}. Frames were mosaiced, and the overscan regions were trimmed. Flat-field frames were used to correct for differences in sensitivity both between detector pixels and across the field. The majority of the cosmic rays were rejected in the individual frames before sky subtraction using a cosmic ray rejection routine. The remainder of the cosmic rays were eliminated using the LACosmic routine (van Dokkum 2001). The sets of 2D spectra were calibrated in wavelength using the arc lamp spectra. Sky emission lines and continuum were removed by averaging the sky spectrum over a number of spatial pixels to reduce the noise level, before subtracting it from all the spatial pixels. Thus the process adds little extra noise to the result. A spectrophotometric standard star (EG 21) was used to correct the measured counts for the combined transmission of the instrument, telescope and atmosphere as a function of wavelength. We reduced the standard star observation with the same instrument configuration as the corresponding scientific data. A 1D spectrum was extracted by adding the central spatial pixels from the standard star observation, and it was used to convert the measured counts from the galaxy spectra into fluxes with erg cm$^{-2}$ s$^{-1}$ \AA{}$^{-1}$ units. The individual science frames were then combined to produce a one final 2D image per galaxy.

The galaxy and associated error spectra were binned in the spatial direction to ensure a minimum signal-to-noise ratio (S/N) of 30 per \AA{} in the H$\beta$ region of the spectrum for measurements as a function of radius. A S/N ratio of 30 per bin was chosen to resolve the optimal number of possible points, whilst still having acceptable errors on the measurements. Thus, the spatial cross-sections are broader with increasing radius from the centre of the galaxy up to a maximum of 30 rows. In all the profiles plotted here, the values of the measurements are plotted at the luminosity-weighted centres of the spatial bins used to derive the parameters. A fifth galaxy (PGC014685) was also observed but with a total exposure time of only 2200 seconds, and more than 30 rows needed to be added to achieve a S/N of 30 per \AA{} at H$\beta$ in the centre of the galaxy. Therefore, this galaxy was eliminated from further analysis. All four other galaxies (and their error spectra) consisted of 8 or more bins, and reach from $\sim0.2 a_{e}$ (2MASXJ17122774-2322108) to $\sim0.9 a_{e}$ (PGC023233). 

The effective half-light radius was calculated as
$a_{\rm e} = \frac{r_{\rm e}(1-\epsilon)}{1-\epsilon \ \mid cos(\mid PA - MA \mid)\mid}$, with $\epsilon$ the ellipticity (data from NED), $r_{\rm e}$ the radius containing half the light of the galaxy (computed from the 2MASS $K$-band 20th magnitude arcsec$^{-2}$ isophotal radius as described in Loubser et al.\ 2008), PA the slit position axis, and MA the major axis. For old stellar populations, these half-light radii do not differ much from those derived using the optical bands (Jarrett et al.\ 2003).

\begin{table*}
\centering
\begin{tabular}{l r r r c r c}
\hline Object  & \multicolumn{1}{c}{Slit} & \multicolumn{1}{c}{Major Axis} & \multicolumn{1}{c}{$r_{\rm e}$} &  $\epsilon$ & \multicolumn{1}{c}{$a_{\rm e}$} & Fraction \\
 & \multicolumn{1}{c}{PA} & \multicolumn{1}{c}{MA} & & & & times \\
 & \multicolumn{1}{c}{(degrees)} & \multicolumn{1}{c}{(degrees)} & \multicolumn{1}{c}{(arcsec)} & & \multicolumn{1}{c}{(arcsec)} & $a_{\rm e}$ \\
\hline ESO541-013 & 197 & 17 & 10.63 & 0.40 & 10.63 & 0.87\\
MCG-02-02-086 & 149 & 149 & 12.10 & 0.22 & 12.10 & 0.71\\
PGC023233 & 193 & 35 & 6.45 & 0.14 & 6.37 & 0.92\\
2MASXJ17122774-2322108 & 135 & 145 & 20.06 & 0.28 & 19.94 & 0.16\\
\hline
\end{tabular}
\caption{Properties of the CCGs. The PA is given as deg E of N. The slit was placed on, or nearby, the major axis of the galaxies. The half-light radii ($r_{\rm e}$) were calculated from the 2MASS catalogue. The last column lists the fraction of the effective half-light radii spanned by the radial profiles measured in this work.}
\label{table:WHTobjects}
\end{table*}

\section{Kinematic measurements}
\label{extinction}
To detect any possible emission line contamination in the CCG absorption line spectra, we use a combination of the \textsc{ppxf} (Cappellari $\&$ Emsellem 2004) and \textsc{gandalf} (Sarzi et al.\ 2006) routines\footnote{We make use of the corresponding \textsc{ppxf} and \textsc{gandalf IDL} (Interactive Data Language) codes which can be retrieved at http:/www.leidenuniv.nl/sauron/.}. \textsc{gandalf} version 1.5 enables a reddening correction to be performed, and encorporates errors. All 985 stars of the MILES stellar library (S\'{a}nchez-Bl\'{a}zquez et al.\ 2006) were used as stellar templates to automatically include $\alpha$-enhancement in the derived optimal template. We followed the procedure described in Sarzi et al.\ (2006), as well as in Loubser $\&$ Soechting (2013). After the kinematics are fixed, a Gaussian template is constructed for each emission line at each iteration, and the best linear combination of both stellar and emission-line templates (with positive weights) is determined. 

None of the four CCGs contained measureable emission lines. We only consider the emission-free absorption line spectra for further analysis.

\textsc{ulyss} (Universit\'{e} de Lyon Spectroscopic analysis Software)\footnote{Available at http://ulyss.univ-lyon1.fr/.} is a full spectrum fitting stellar population synthesis code which can be used to determine the stellar atmospheric parameters, star formation and metal enrichment histories of galaxies (Koleva et al.\ 2009). The entire observed spectrum of an object is fitted against a model which is expressed in the form of a linear combination of non-linear components. These components are given in the form of non-linear functions of ages, metallicities ([Fe/H]), wavelengths and other parameters. Following Bouchard et al. (2010), the models are combined with a line-of-sight velocity distribution (LOSVD) and multiplied with a polynomial to absorb the effects of flux calibration errors and the Galactic extinction which may influence the shape of the spectra (see Viljoen $\&$ Loubser 2013 for a detailed demonstration of this method). The \textsc{ulyss} software package was chosen to analyze the SFHs of the CCGs because this software has features that enable the user to better understand the structure of the parametre space by constructing $\chi^{2}$-maps which determines the degeneracies and the errors on the parameters (see Koleva et al. 2008).

The Pegase.HR model with the ELODIE v3.1 stellar library was used (Prugniel et al. 2007). This library covers a wavelength range of 3892 -- 6800 \AA{} and allows the components of the models to be expanded by defining the type of initial mass function (IMF), evolutionary track and star formation rate. The evolutionary tracks of the isochrones are then computed using Padova 1994 (Bertelli et al.\ 1994). Du et al.\ (2010) defined these isochrones as being solar scaled at various values of the total metallicity, Z. This version of the stellar library computes the SSPs with the Salpeter IMF (Salpeter 1955) with a mass of 0.1 $\leq$ M$_\odot \geq$ 120 and a slope of 1.35. This model covers an age range of 0.01; 20.00 Gyr and a [Fe/H] range of --2.30; 0.69 dex. We used a new calibration of this model which includes [Mg/Fe] = 0. Variable $\alpha-$enhancements models are still to be incorporated into the \textsc{ulyss} software.

The spatially resolved velocity and velocity dispersion gradients are shown in Figures \ref{fig:velocity} and \ref{fig:dispersion}. In all the spatially resolved profiles plotted here, the values of the parameters are plotted at the luminosity-weighted centres of the spatial bins used to derive the parameters. The spatially resolved kinematics was then used to fit spatially-resolved, SSP-equivalent ages and metallicities. For the purpose of this study, the SFHs of the galaxies were analyzed by fitting an SSP and composite stellar populations (CSPs) against the observed spectra of the galaxies. The errors given are the standard deviation (1$\sigma$) on the average values of the ages and [Fe/H] (where [Mg/Fe]=0). These errors are determined by the mpfit\footnote{http://cow.physics.wisc.edu/$\sim$craigm/idl/idl.html} algorithm which uses a covariance matrix to determine the best fit and the 1$\sigma$ errors on the parameters (Koleva et al. 2009).
 
It is sometimes found that for a given galaxy, an SSP model can not provide a satisfactory fit to the observed spectrum of the galaxy, implying that the galaxy experienced more than one SF epoch and hence CSPs are used to represent the SFH. Each galaxy was fitted against three components (a young, intermediate and old component) and the optimal solutions were given in terms of the age, [Fe/H], light fraction (LF), mass fraction (MF) and the errors on these values, following the procedure described below. 

The time (age) axis can be divided into intervals by setting limits. For this paper three or less epochs were assumed: an older, an intermediate and a young stellar population. The three epochs were defined by setting limits on the ages (similar to the procedures in Du et al.\ 2010 and Bouchard et al.\ 2010), while no limits were imposed on [Fe/H]. The age boundaries were (i) 12 to 20 Gyr; (ii) 4 to 12 Gyr and; (iii) 0.01 to 1.00 Gyr. Whenever CSP fits delivered results close to the boundaries of at least one of the three boxes, the upper or lower age boundaries of the boxes were changed until the results converged (in order not to artificially limit the three components to these chosen values --- the three chosen intervals were just used as initial guesses). This method is also justified by the fact that whenever four or more components were fitted the weights of the additional components were zero. Thus there is no dependence of the young population mass fraction on any input parameter. Note that the oldest ages in these galaxies quite often hit the upper limits of the models, which are older than the current age of the Universe (see discussion in S\'{a}nchez-Bl\'{a}zquez et al.\ 2009). Therefore, all interpretations of these ages should be based on relative differences in ages which are much more reliable than absolute values.

The constructed $\chi^{2}$-maps are used to visualize the degeneracies between the parameters and to reveal the presence of the local minima. The $\chi^{2}$-maps are based on a grid of initial guesses and a global minimization is then performed to evaluate the region where the parameter space converges to the absolute minimum of the $\chi^{2}$ (Koleva et al.\ 2009). Figures \ref{fig:ESO_chi} to \ref{fig:2MAS_chi} show the young (or only) stellar component derived for each of the four CCGs. The radial extent and percentage light fractions of the young stellar populations are shown in Figure \ref{fig:Fractions}.

\begin{figure*}
   \centering
\includegraphics[scale=0.7]{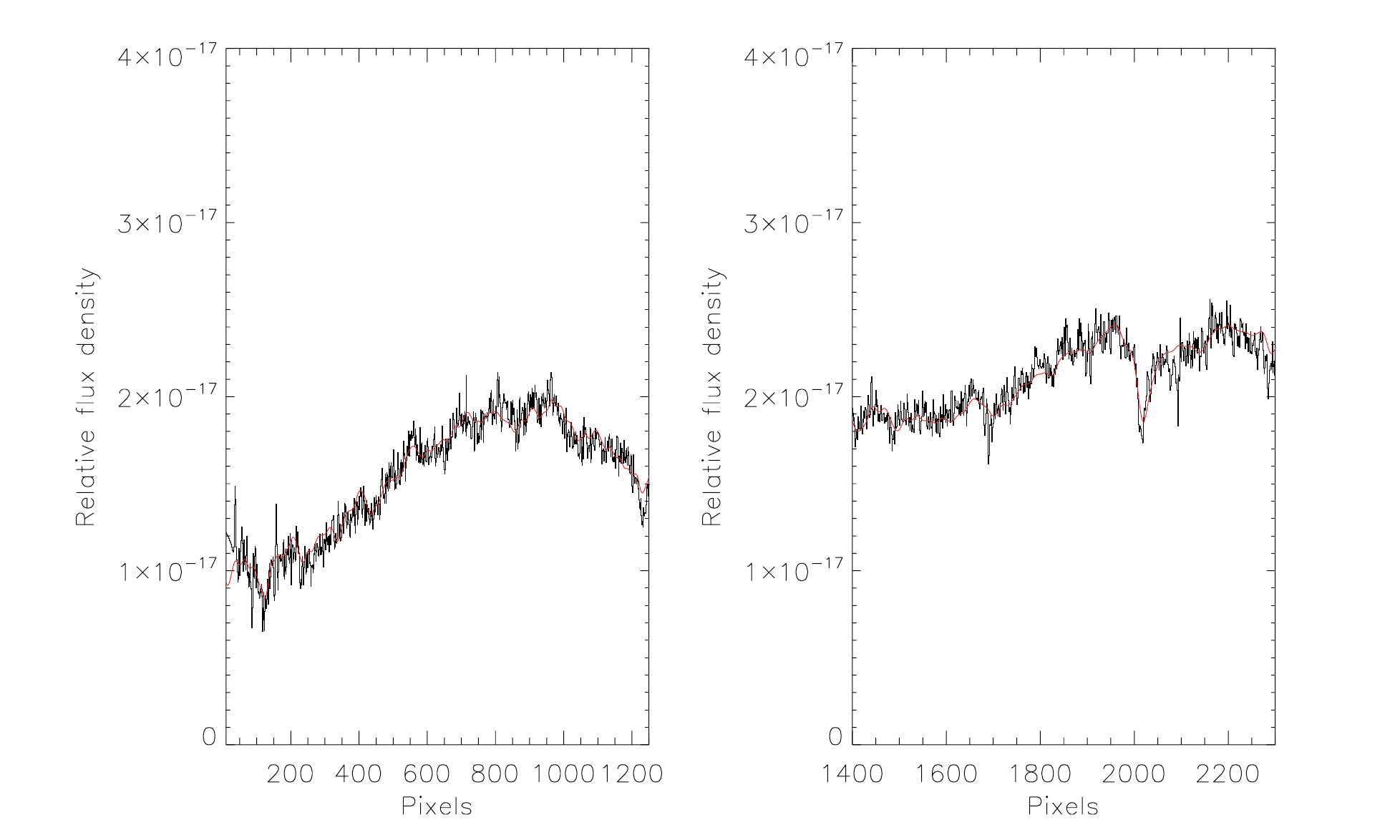}
\caption{\textsc{gandalf} fits to the central spectra of MCG-02-02-086 to look for possible weak emission. The grey line indicates the combination of stellar spectra from the MILES library that delivered best fit. The panels represents the CCDs.}
\label{fig:Fits} 
\end{figure*}

\begin{figure*}
   \centering
   \includegraphics[scale=0.6]{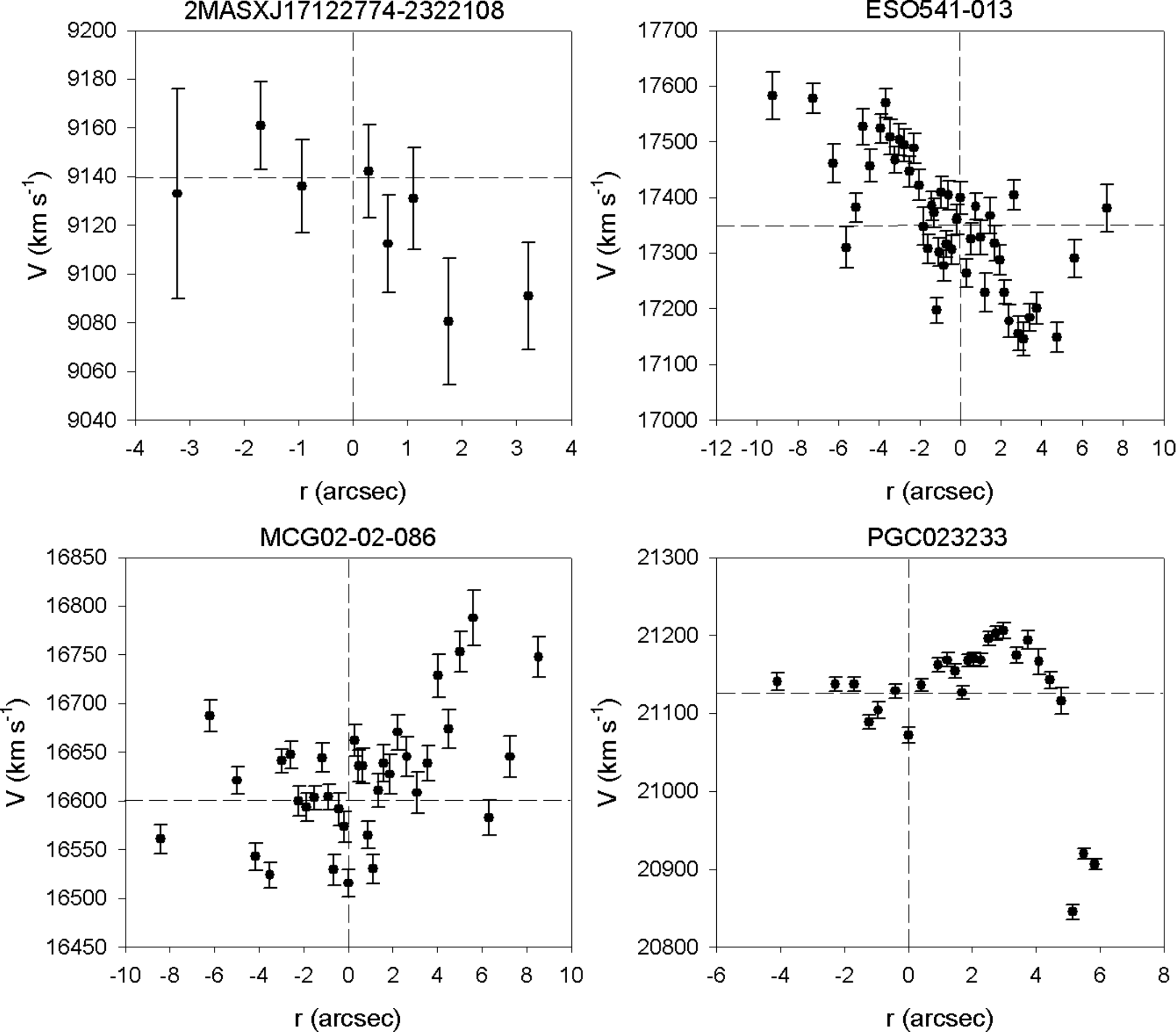}
   \caption{Spatially-resolved radial velocities of the four CCGs.}
   \label{fig:velocity}
\end{figure*}

\begin{figure*}
   \centering
   \includegraphics[scale=0.6]{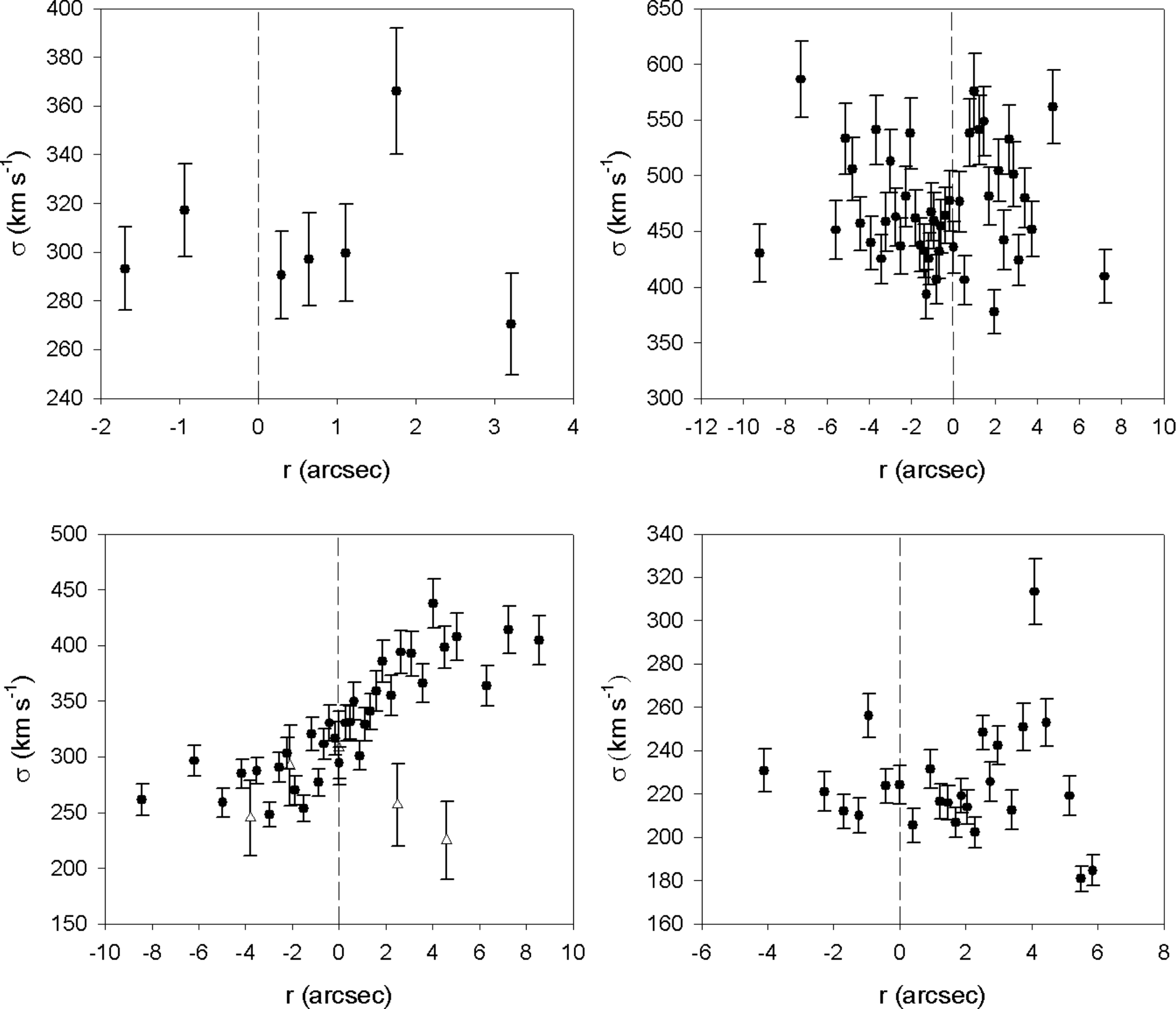}
   \caption{Spatially-resolved velocity dispersion measurements of the four CCGs. From left to right, top to bottom: 2MASXJ17122774-2322108, ESO541-013, MCG02-02-086, PGC023233. The empty triangles in MCG02-02-086 is data from Fisher et al.\ (1995) also taken along the major axis.}
   \label{fig:dispersion}
\end{figure*}

\begin{figure*}
   \centering
 \mbox{\subfigure[ESO541-013]{\includegraphics[scale=1]{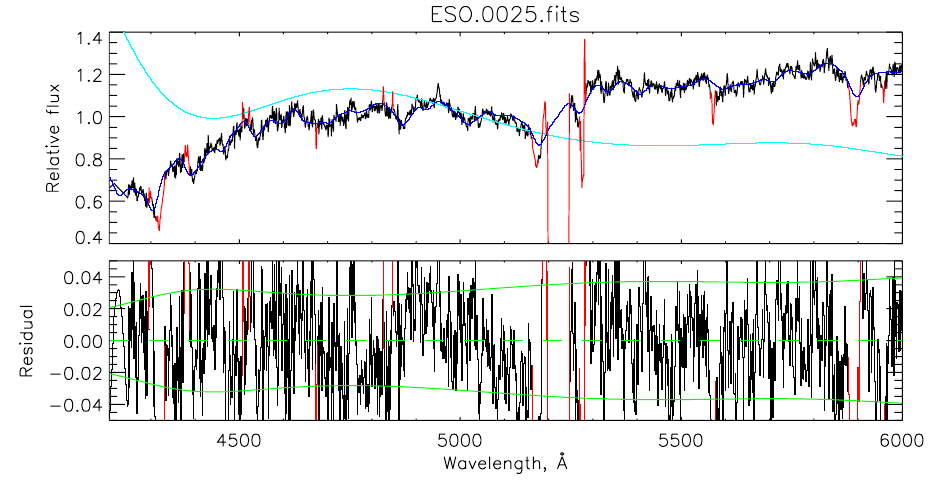}}\quad
         \subfigure[MCG-02-02-086]{\includegraphics[scale=1]{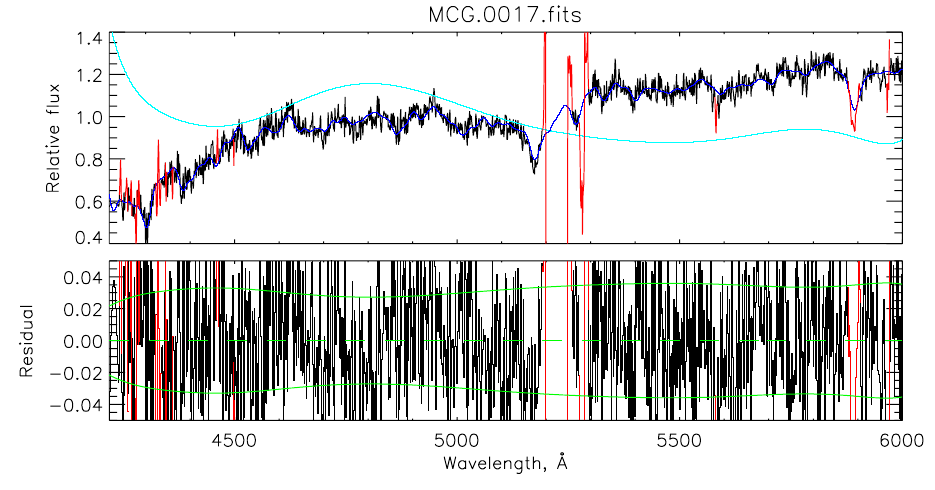}}}
\mbox{\subfigure[PGC023233]{\includegraphics[scale=1]{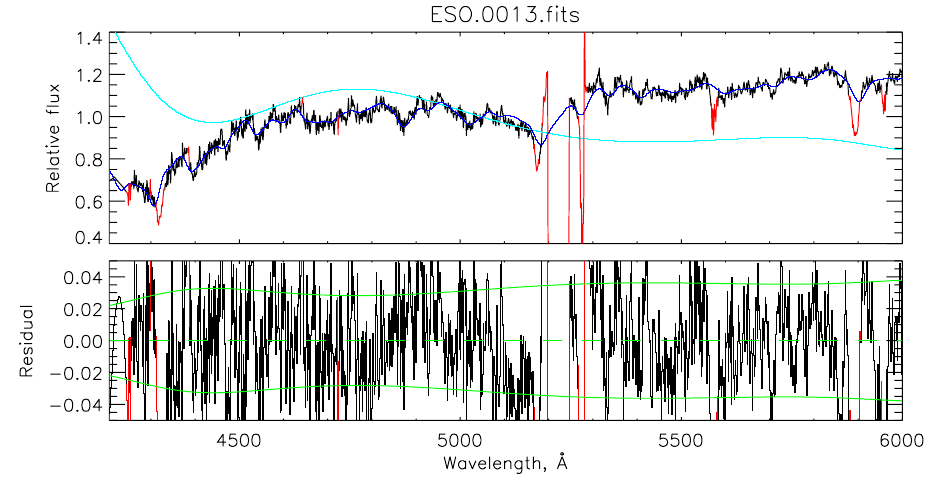}}\quad
\subfigure[2MASXJ17122774-2322108]{\includegraphics[scale=1]{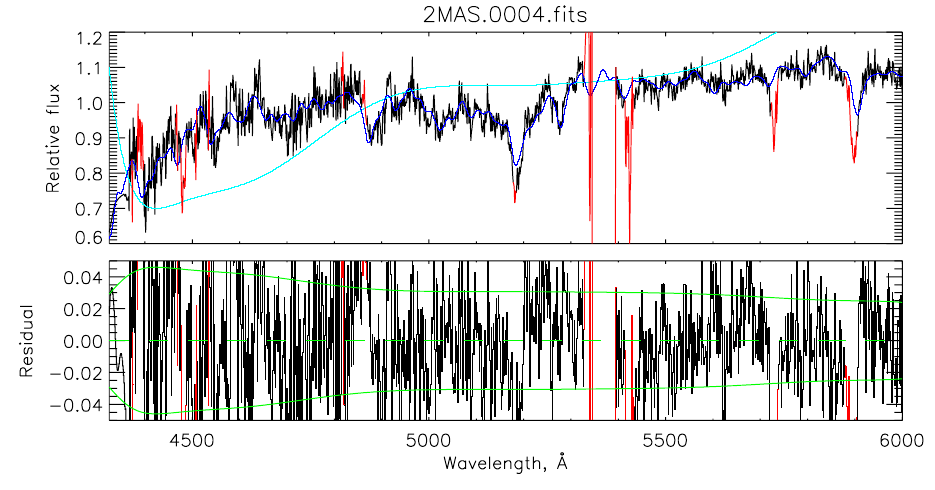}}}
\caption{\textsc{ulyss} fits to the central spectra of the four galaxies to derive the kinematics and stellar population properties. The red regions were masked in the fit.}
\label{fig:FitsUlyss} 
\end{figure*}

\begin{figure}
   \centering
   \includegraphics[scale=0.5]{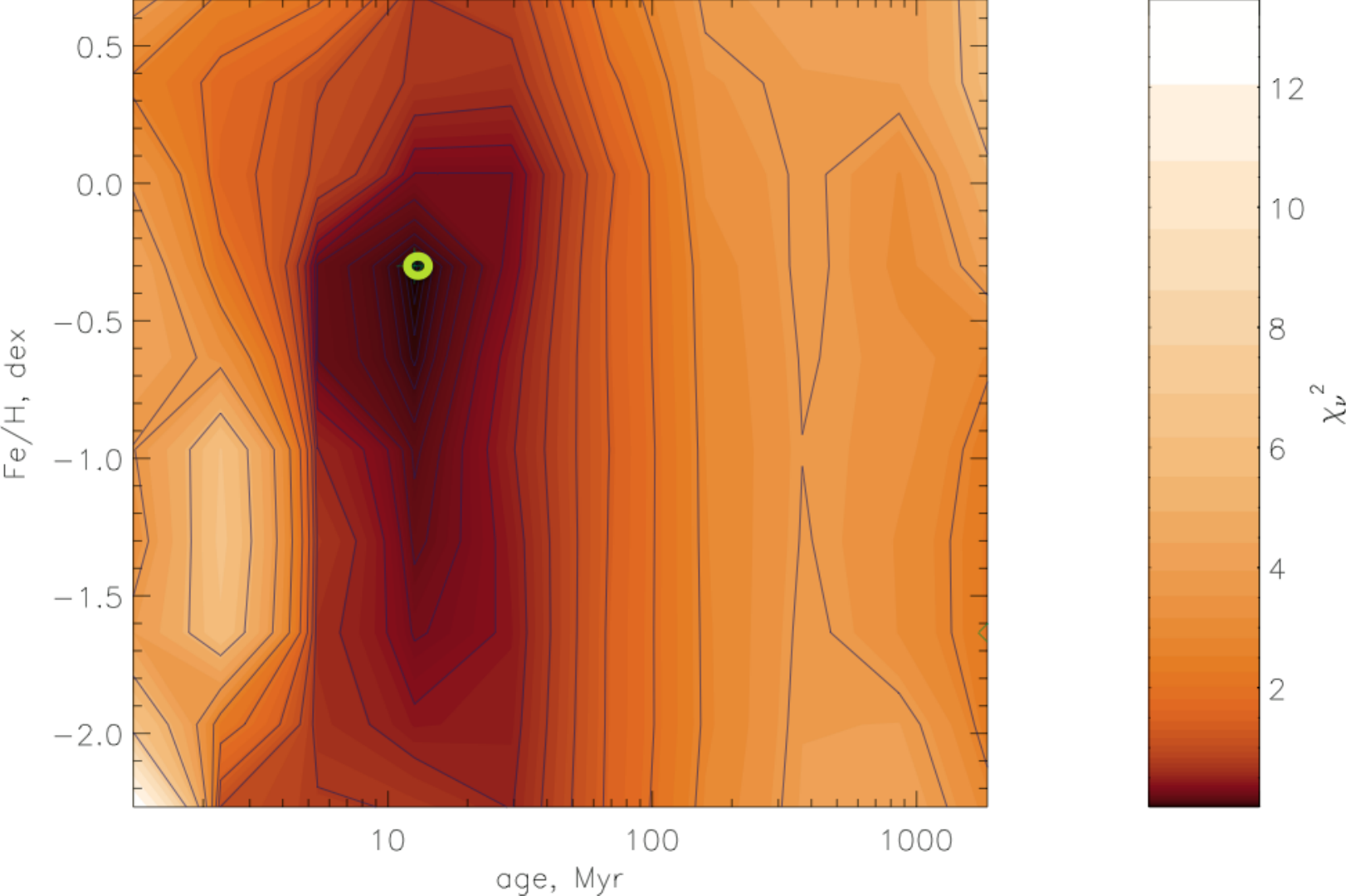}
   \caption{$\chi^{2}$-map of the young component of ESO541-013. The global minimum is indicated with a green circle.}
   \label{fig:ESO_chi}
\end{figure}

\begin{figure}
   \centering
   \includegraphics[scale=0.5]{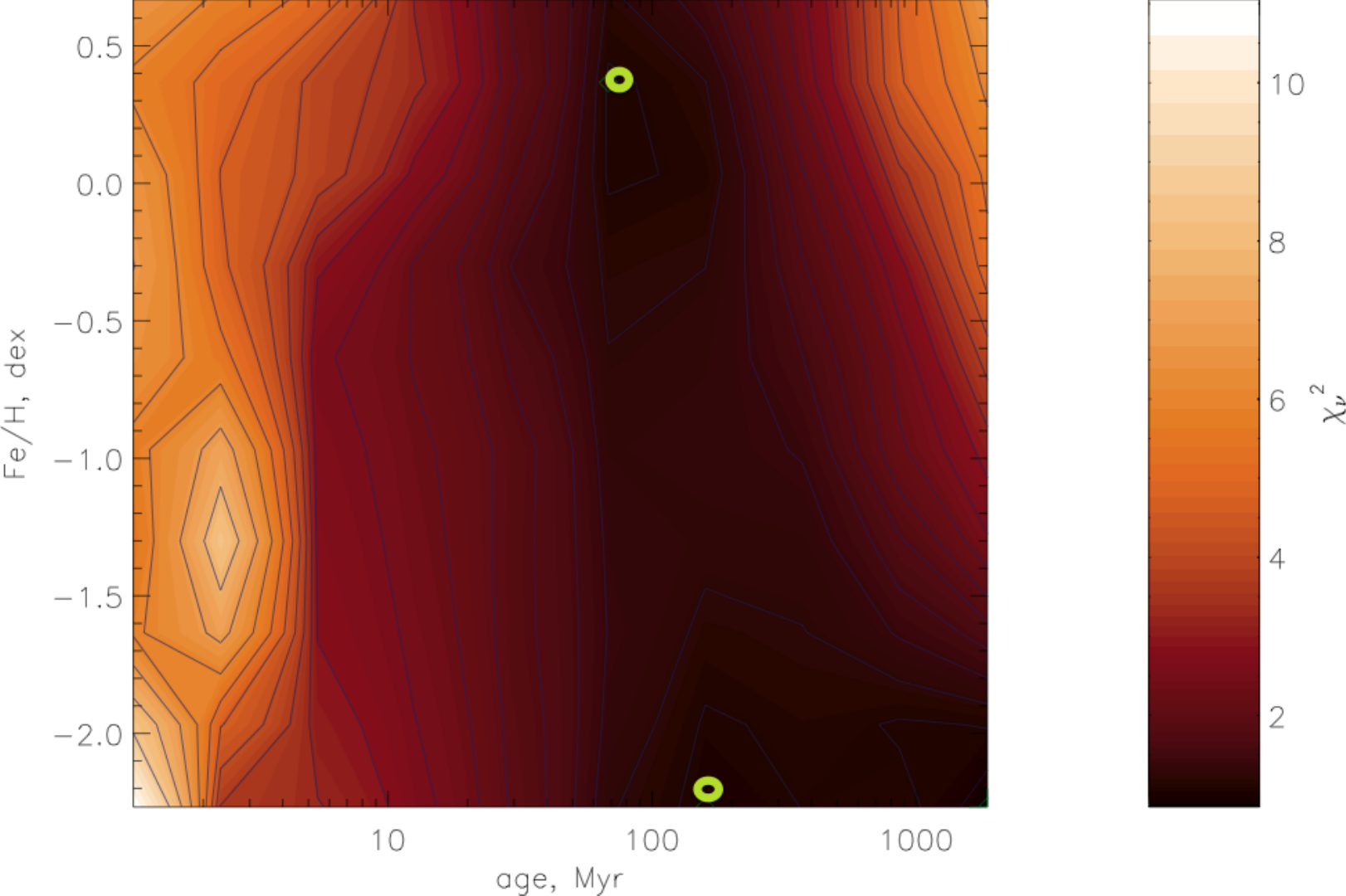}
   \caption{$\chi^{2}$-map of the young component of MCG-02-02-086. The global minimum can be seen as a green circle at the bottom (right corner) of the map, and the next probable solution is also shown at the top (middle) of the map.}
   \label{fig:MCG_chi}
\end{figure}

\begin{figure}
   \centering
   \includegraphics[scale=0.5]{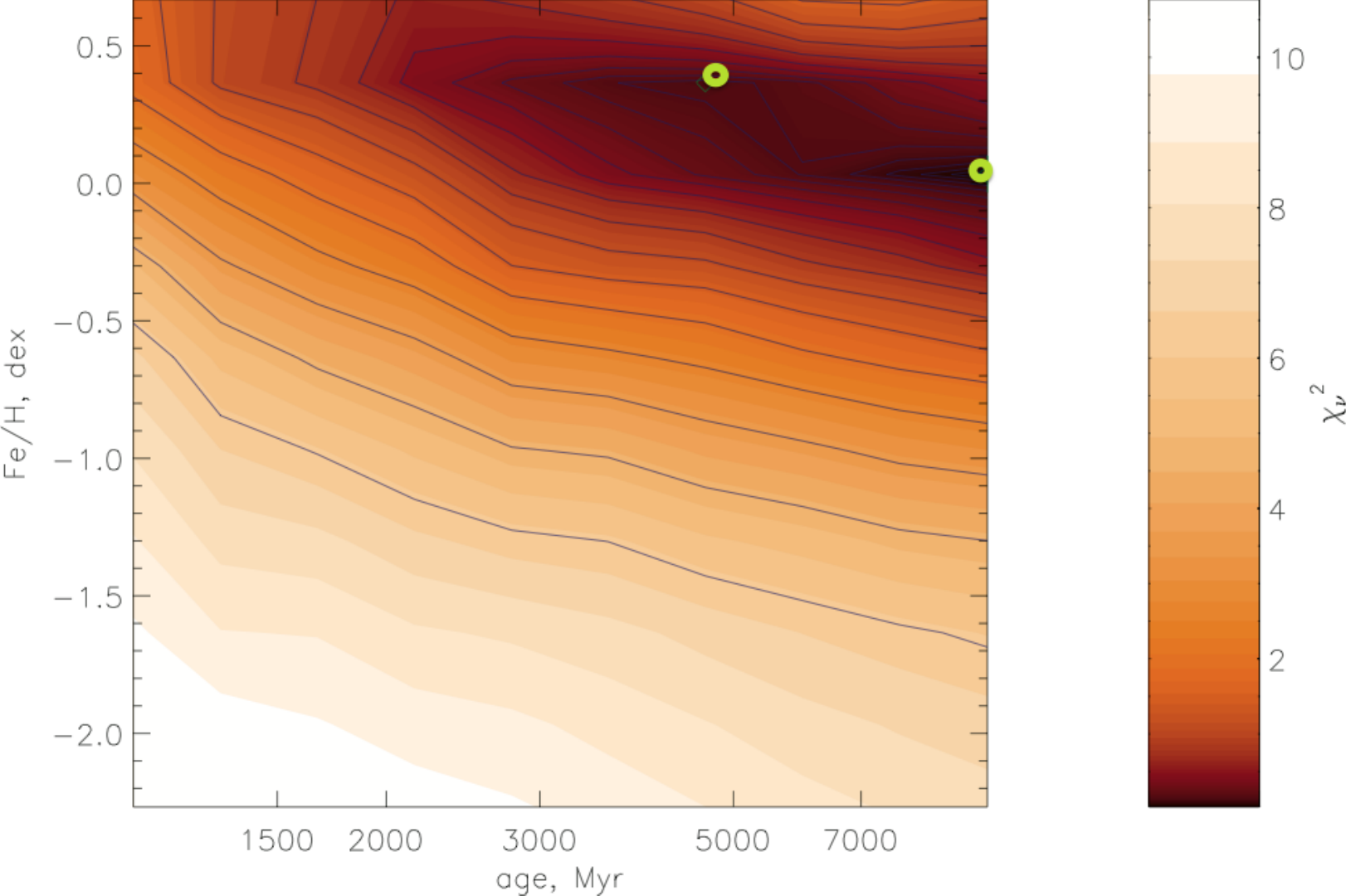}
   \caption{$\chi^{2}$-map of the only component of PGC023233 (SSP is the best fit). The global minimum can be seen as a green circle at the right of the map, with the second most probable solution to the left.}
   \label{fig:PGC_chi}
\end{figure}

\begin{figure}
   \centering
   \includegraphics[scale=0.5]{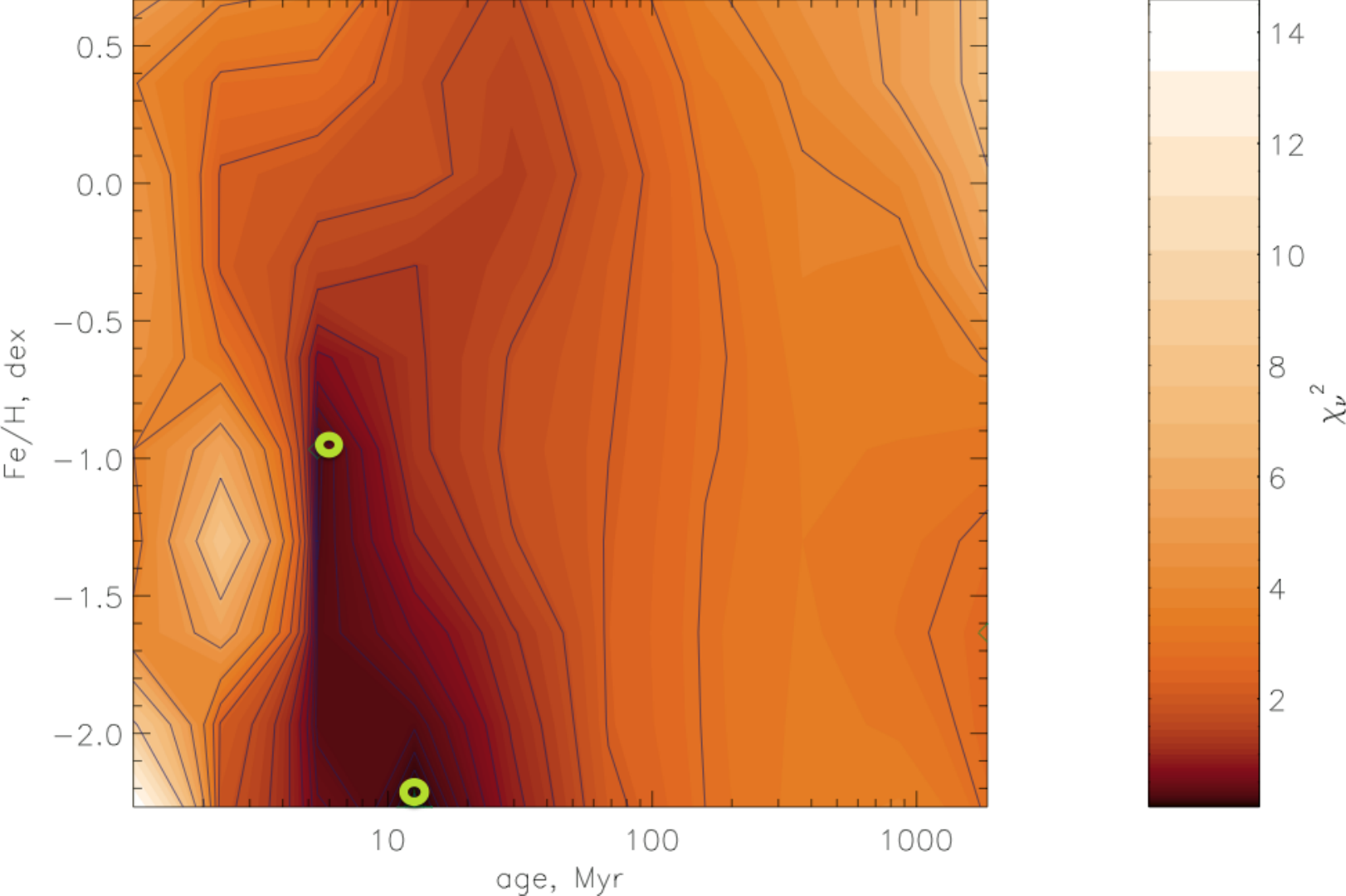}
   \caption{$\chi^{2}$-map of the young component in the centre of 2MASXJ17122774-2322108. The global minimum can be seen as a green circle at the bottom of the map, with the second most probable solution above.}
   \label{fig:2MAS_chi}
\end{figure}

\begin{figure*}
   \centering
 \mbox{\subfigure[ESO541-013]{\includegraphics[scale=0.4]{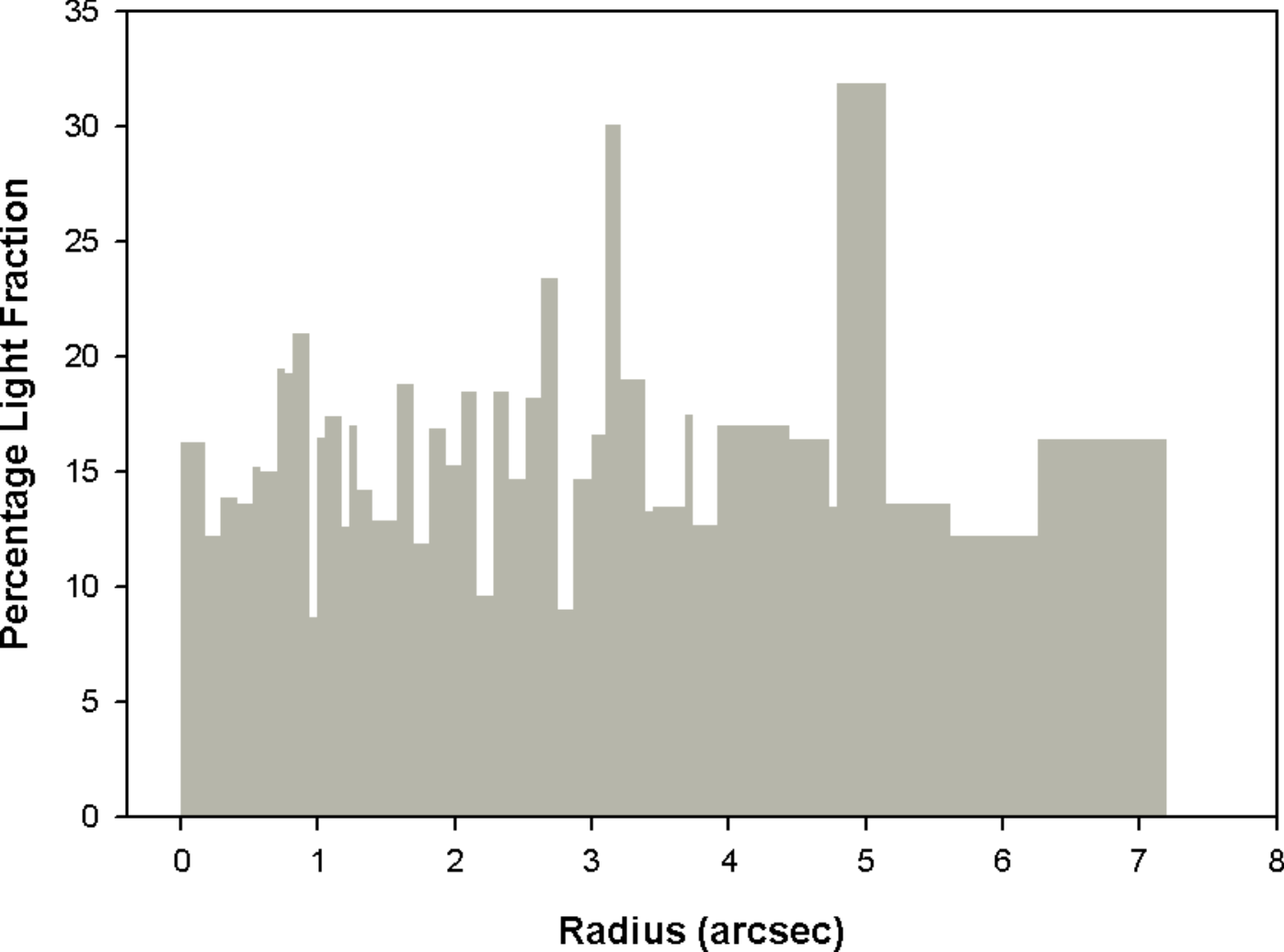}}\quad
         \subfigure[MCG02-02-086]{\includegraphics[scale=0.4]{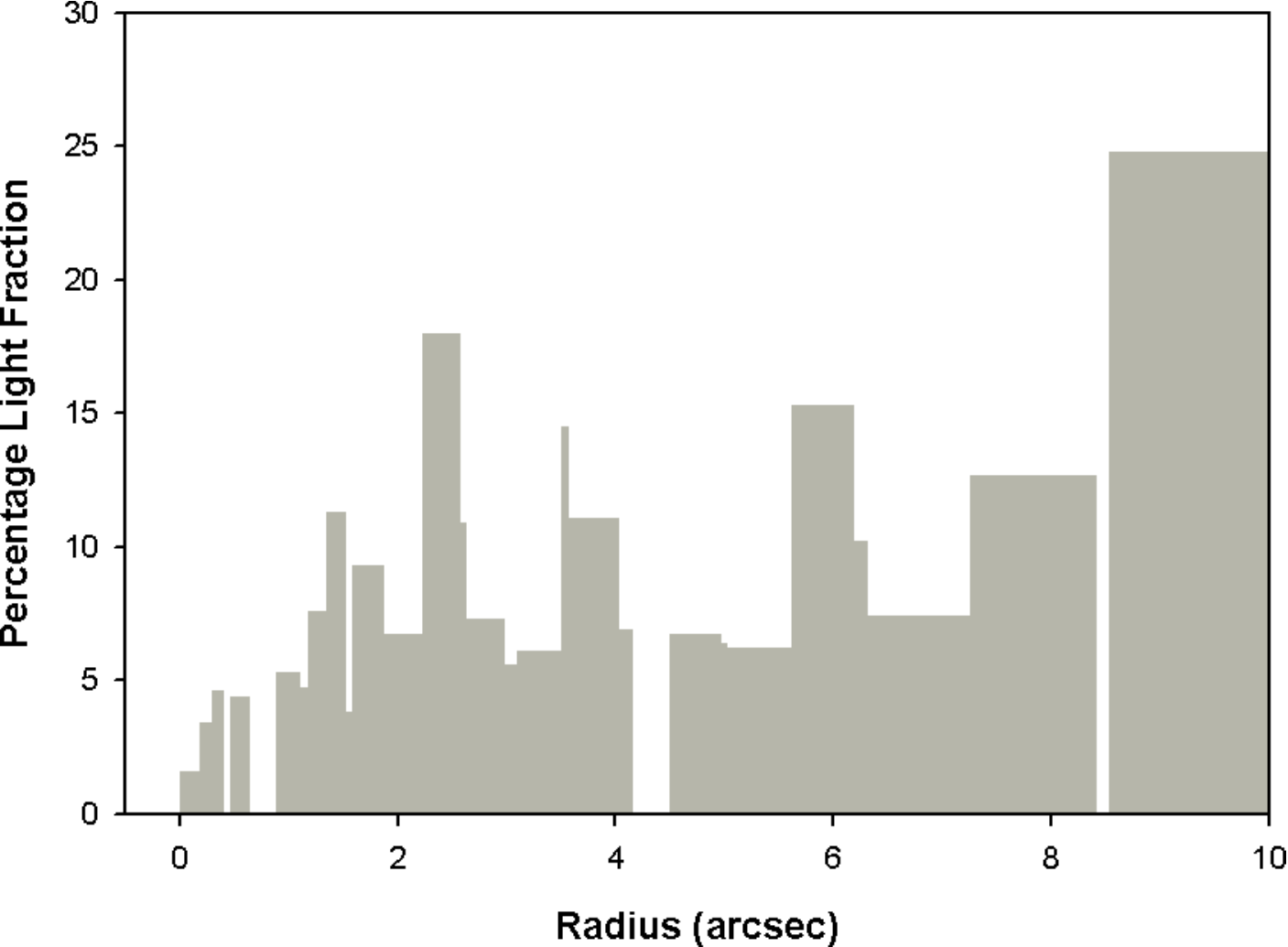}}}
\mbox{\subfigure[PGC023233]{\includegraphics[scale=0.4]{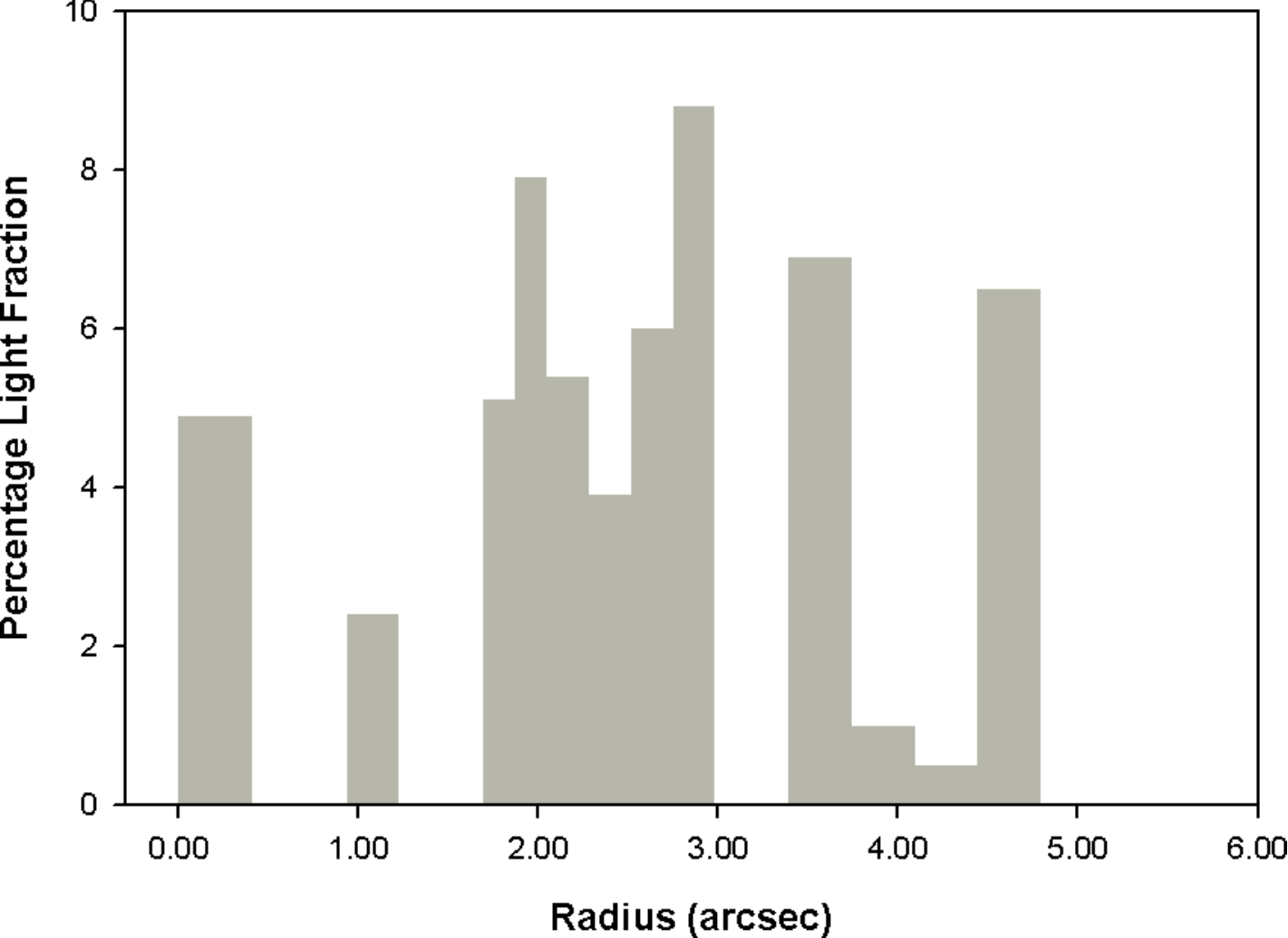}}\quad
\subfigure[2MASXJ17122774-2322108]{\includegraphics[scale=0.4]{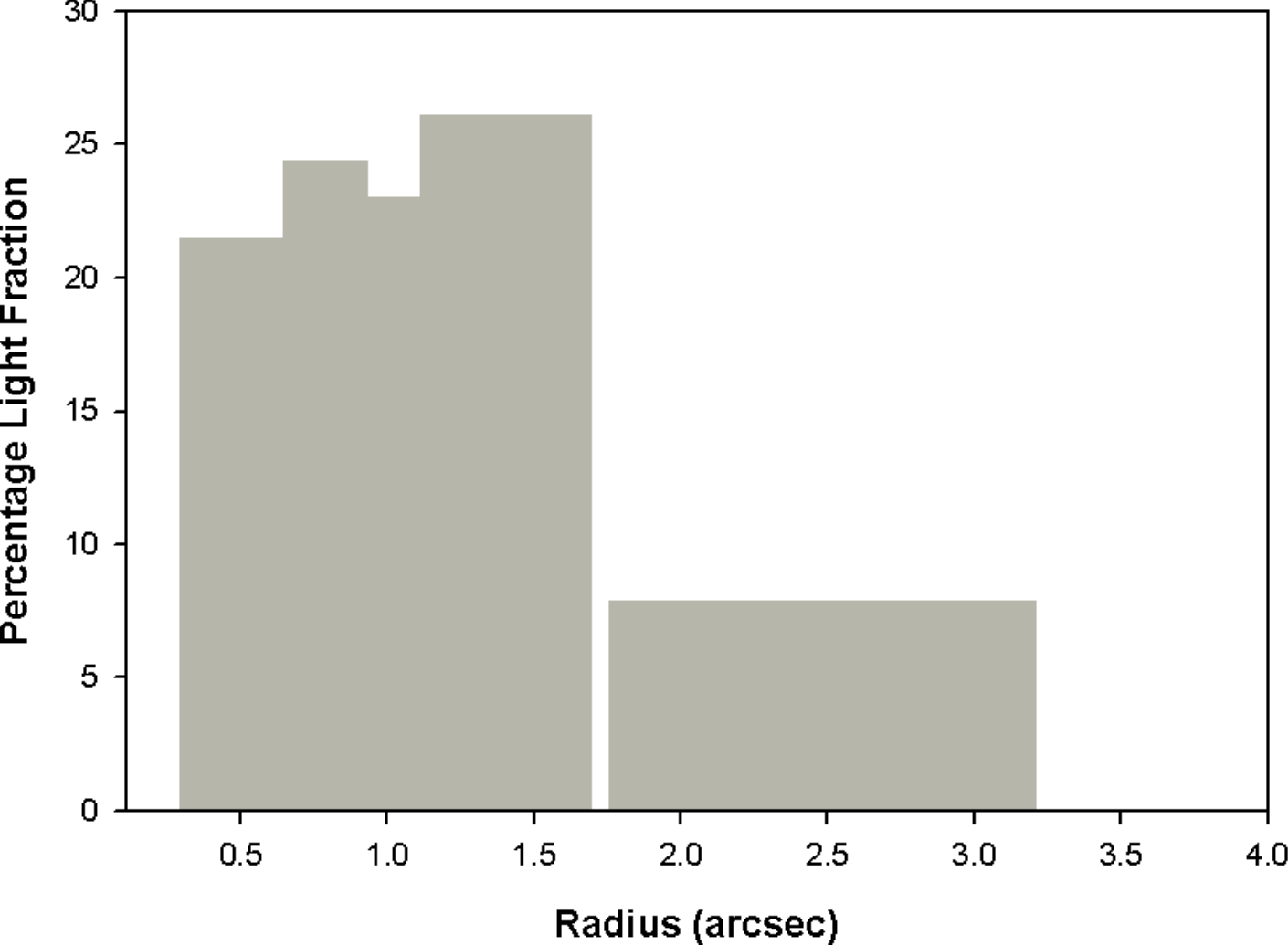}}}
\caption{The radial extent and percentage light fractions of the young stellar populations.}
\label{fig:Fractions} 
\end{figure*}

\section{Individual galaxies}
\label{figures_individual}

\subsection{ESO541-013 -- Abell 0133}

White et al.\ (1997) classified this cluster as a cooling-flow cluster, however B\^{\i}rzan et al.\ (2004) and McCarthy et al.\ (2004) classified it as a non-cooling flow cluster. More recently,  McDonald et al.\ (2010) fit a null spectroscopic mass deposition rate to this cluster, whilst according to Hudson et al.\ (2010), Abell 0133 is a strong cool core cluster (based on the cooling time, which, in their analysis, seems to be the most robust indicator of the cool core strength). The classification of a host cluster as a cooling vs.\ a non-cooling core is troublesome, so for our purposes we will just refer to this cluster as having a zero spectroscopically determined mass deposition rate without making further assumptions. The CCG showed extended H$\alpha$ emission in McDonald et al.\ (2010). A single thin H$\alpha$ filament (coincident with X-ray) extends northeast from the centre of the cluster for $\sim$ 25 kpc.

The $\chi^{2}$-values of the SSP and CSP fits do not differ significantly, but the multiple component fitting was found to be more consistent (with the properties and number of components derived for the various spatial bins agreeing within the errors)\footnote{We do not expect dramatic stellar population (in particular age) gradients for these galaxies, see Loubser $\&$ S\'{a}nchez-Blazquez (2012).}, and showed that the majority of the bins consisted of two components: a large, very old component (bordering the upper limit of the stellar populations model at 20 Gyr), and a smaller very young component ($<$ 100 Myr) contributing up to $\sim$ 20 per cent of the light. 

\subsection{MCG-02-02-086 -- Abell 0085}

This CCG is hosted by a cooling flow cluster (see Table \ref{table:objects}), but lacks a cD envelope (as summarised in Fisher, Illingworth $\&$ Franx 1995). It showed nuclear H$\alpha$ emission in McDonald et al.\ (2010), and a large ring can be seen in the X-ray image (but without an H$\alpha$ counterpart), resembling the outer edge of a bubble (McDonald et al.\ 2010). Based on the high H$\alpha$/NUV ratio, McDonald et al.\ (2010) postulated that the gas could be ionised by a recent burst of star formation in the CCG centre.

Similarly to ESO541-013, the $\chi^{2}$-values of the SSP and CSP fits do not differ significantly, but the multiple component fitting was found to be more consistent. It showed that the majority of the bins consisted of two components, a large, very old component (bordering the upper limit of the stellar populations model at 20 Gyr), and a very small young component (of less than 1 Gyr contributing $\sim$ 5 to 10 per cent of the light). This agrees with the scenario McDonald et al.\ (2010) postulated above. 

\subsection{PGC023233 -- Abell 0644}

This host cluster is classified as a cooling-flow cluster (see Table \ref{table:objects}), and exhibited nuclear H$\alpha$ emission in McDonald et al.\ (2010).

For this galaxy, the CSP fitting resulted in an inconsistent number of components (several bins consisting of 1,2 or 3 components -- see Table \ref{table:CSP_PGC}), and the SSP fitting delivered an age estimate ranging between 6000 to 20 000 Myr (the upper limit of the model), with the majority of the bins being of intermediate age. Thus for this galaxy, it is difficult to accurately pinpoint the most probable age estimate and number of components. 

It can be seen in figure 6 in McDonald et al.\ (2010) that the upper limit of the global NUV luminosity of Abell 0644 (measured from archival GALEX images) lies barely above the star formation law (Kennicutt 1998). It is therefore not obviously younger or with more star formation activity than the other three galaxies analysed in this article.

\subsection{2MASXJ17122774-2322108 -- Ophiuchus}

This low density cluster was classified as a cooling-flow cluster (White et al.\ 1997), whereas McDonald et al.\ (2010) fit a null spectroscopic mass deposition rate to this cluster,
The CCG showed nuclear H$\alpha$ emission in McDonald et al.\ (2010).

Tables \ref{table:SSP_2MAS} and \ref{table:CSP_2MAS} shows the SSP and CSP equivalent ages and metallicities derived for the galaxy. The $\chi^{2}$-values of the SSP and CSP fits do not differ significantly, but once again the multiple component fitting was found to be more consistent, and showed that the majority of the bins consisted of a larger intermediate component of $\sim$ 7000 Myr and a smaller very young component of $\sim$ 10 Myr which contributes up to 25 per cent of the light. 


\subsection{Discussion}

It is interesting to ask whether any particular distribution of star formation, and hence the young stars, can be detected in the observations. Figure \ref{fig:Fractions} shows the radial distribution of the young components. Only one galaxy (2MASXJ17122774-2322108) shows a decrease in the fraction of the young stars as the radius increase from the centre of the galaxy. The other galaxies show that the young stars are equally present throughout the radius that our observations cover.

It is further necessary to take note of any possible AGN activity in the central galaxies, as the weak emission lines from AGN can fill in some of the absorption lines that is important for stellar population age measurements (such as the Balmer lines), and can ultimately lead to older derived age estimates. The total 1.4 GHz radio flux at the centre of a cluster may be indicative of AGN activity. For the four clusters we study here, we used the 1.4 GHz fluxes measured from the NRAO VLA Sky Survey (NVSS; Condon et al.\ 1998). All but one galaxy (PGC023233) show radio emission in Figures \ref{fig:ESO_NVSS} to \ref{fig:2MAS_NVSS}. The H$\alpha$ filaments measured in McDonald et al.\ (2010) are very small ($<$ 4 kpc) in comparison with the extensive radio emission, but are located in the centres of the 1.4 GHz and X-ray emission. 

Previous optical emission-line and ultraviolet studies can then be included to quantitatively determine whether AGN contamination is present in the optical spectra. Three of the four galaxies (all except ESO541-013) have nuclear H$\alpha$ emission. ESO541-013 is classified as a non-emission line galaxy by Edwards et al.\ (2007), as its H$\beta$ equivalent width is --0.08 $\pm$ 0.09 \AA{} (after correcting for the underlying absorption), whereas emission galaxies are classified with H$\beta$ equivalent width as $>$ 0.5 \AA{}. Using this same criteria, MCG-02-02-086 is classified as a galaxy with star forming activity as its H$\beta$ equivalent width is 1.13 $\pm$ 0.21 \AA{} (and no AGN activity is suspected). From the GALEX NUV excess, Hicks, Mushotzky $\&$ Donahue (2010) estimate a star formation rate of 0.004 $M_{\odot}$/yr (assuming continuous star formation over 20 Myr) for PGC023233 (with no evidence of AGN activity). 2MASXJ17122774-2322108, however, is a suspected AGN (see Murgia et al.\ 2010 and Perez-Torres et al.\ 2009). It therefore seems that for three of the four galaxies, no AGN are hosted but weak star formation for MCG-02-02-086 and PGC023233 are confirmed by previous observations. No emission features were found within our slit positions for all four galaxies.

No significant stellar population gradients are measured along the major axis, which are mostly aligned with the radio emission (Figure \ref{fig:Thumbnails} compared to Figures \ref{fig:ESO_NVSS} to \ref{fig:2MAS_NVSS}). As discussed in Loubser $\&$ Sanchez-Blazquez (2012), because the models predict that the majority of mergers happen at recent times, the gas content of the accreted galaxies is believed to be low (e.g.\ Dubinski 1998; Conroy, Wechsler $\&$ Kravtsov 2007; De Lucia $\&$ Blaizot 2007) and these mergers would therefore not change the central ages and metallicities of the CCGs. This would explain the lack of large differences between the stellar populations of CCGs and normal galaxies. However, these merger or accretion events are expected to change stellar population gradients, as dry minor mergers would deposit metal poor stars outwards (Kawata et al.\ 2006). Thus, metallicity gradients in galaxies affected by mergers would be steeper than their non-merger counterparts. We can therefore not deduce any merger events from these observations which were centered on the nucleus of these extended galaxies.

\begin{table*}
\centering
\begin{tabular}{r c r@{$\pm$}l r@{$\pm$}l}
\hline Radius  & \multicolumn{1}{c}{$\chi^{2}$} & \multicolumn{2}{c}{Age} & \multicolumn{2}{c}{[Fe/H]} \\
(arcsec) & \multicolumn{1}{c}{} & \multicolumn{2}{c}{(Myr)} & \multicolumn{2}{c}{} \\
\hline --3.22&1.599&9796&2714&--0.556&0.094\\
--1.70&1.544&4992&808&--0.032&0.054\\
--0.94&1.032&4519&744&0.082&0.046\\
0.29&1.136&4196&817&--0.010&0.054\\
0.64&0.087&5062&829&--0.059&0.054\\
1.11&1.081&7178&1316&--0.171&0.062\\
1.76&1.325&14546&3415&--0.423&0.076\\
3.22&2.202&9892&2139&--0.463&0.076\\
\hline
\end{tabular}
\caption{SSP-equivalent ages and metallicities fitted to 2MASXJ17122774-2322108.}
\label{table:SSP_2MAS}
\end{table*}

\begin{table*}
\centering
\begin{tabular}{r c r@{$\pm$}l r@{$\pm$}l}
\hline Radius  & \multicolumn{1}{c}{$\chi^{2}$} & \multicolumn{2}{c}{Age} & \multicolumn{2}{c}{[Fe/H]} \\
(arcsec) & \multicolumn{1}{c}{} & \multicolumn{2}{c}{(Myr)} & \multicolumn{2}{c}{} \\
\hline --9.24 & 1.901& 3977&837&0.282&0.031\\
--7.25 & 1.294&13700&2345&--0.093&0.056\\
--6.26 & 1.303&15464&2572&--0.179&0.044\\
--5.62 & 1.219&12303&2091&0.006&0.046\\
--5.15 & 1.220&5855&427&0.197&0.038\\
--4.80 & 1.289&13543&1962&--0.066&0.049\\
--4.45 & 0.877&10870&1491&--0.009&0.053\\
--3.92 & 0.999&16376&3123&0.086&0.041\\
--3.69 & 1.090&6472&267&0.166&0.033\\
--3.45 & 0.929&5781&509&0.154&0.035\\
--3.22 & 0.926&6262&369&0.120&0.037\\
--3.00 & 0.829&12895&2072&0.044&0.040\\
--2.75 & 0.799&10347&1656&--0.001&0.046\\
--2.52 & 0.849&5794&419&0.277&0.036\\
--2.28 & 0.713&15345&2223&0.003&0.038\\
--2.05 & 0.700&6198&537&0.073&0.035\\
--1.81 & 0.684&6196&368&0.117&0.036\\
--1.58 & 0.619&12679&1946&--0.043&0.051\\
--1.40 & 0.647&15853&2568&0.012&0.039\\
--1.29 & 0.835&11246&1676&0.015&0.044\\
--1.17 & 0.799&12359&2019&0.010&0.043\\
--1.05 & 0.604&14442&2255&--0.001&0.040\\
--0.94 & 0.827&6047&352&0.133&0.035\\
--0.82 & 0.913&12566&2196&0.032&0.042\\
--0.70 & 0.745&15285&2239&0.000&0.040\\
--0.58 & 0.647&4072&653&0.192&0.026\\
--0.41 & 0.716&6395&435&0.095&0.035\\
--0.18 & 0.746&3487&242&0.286&0.046\\
0.00 & 0.624&6034&268&0.192&0.035\\
0.29 & 0.614&5961&417&0.124&0.036\\
0.53 & 0.612&6272&247&0.180&0.033\\
0.76 & 0.623&16075&2647&--0.001&0.038\\
0.99 & 0.707&10454&1642&--0.001&0.045\\
1.23 & 0.705&6007&291&0.194&0.035\\
1.46 & 0.749&13429&2121&--0.045&0.053\\
1.70 & 0.844&15986&2738&--0.035&0.048\\
1.93 & 0.846&15614&2544&0.015&0.042\\
2.16 & 0.906&16122&2735&--0.007&0.043\\
2.40 & 0.880&17135&3877&--0.002&0.033\\
2.63 & 0.970&6361&306&0.144&0.035\\
2.87 & 0.975&11949&1628&--0.095&0.054\\
3.10 & 0.968&15958&2818&0.023&0.040\\
3.39 & 0.915&6057&416&0.121&0.038\\
3.74 & 0.929&20000&1000&0.001&0.023\\
4.74 & 1.052&6160&399&0.108&0.036\\
5.62 & 1.110&11960&1875&--0.002&0.047\\
7.20 & 1.381&11110&1831&--0.002&0.054\\
\hline
\end{tabular}
\caption{SSP-equivalent ages and metallicities fitted to ESO541-013.}
\label{table:SSP_ESO}
\end{table*}

\begin{table*}
\centering
\begin{tabular}{r c r@{$\pm$}l r@{$\pm$}l}
\hline Radius  & \multicolumn{1}{c}{$\chi^{2}$} & \multicolumn{2}{c}{Age} & \multicolumn{2}{c}{[Fe/H]} \\
(arcsec) & \multicolumn{1}{c}{} & \multicolumn{2}{c}{(Myr)} & \multicolumn{2}{c}{} \\
\hline--8.42 & 2.476&18507&2541&--0.226&0.034\\
--6.20&2.368&20000&1000&--0.056&0.025\\
--4.97&2.244&20000&1000&--0.123&0.024\\
--4.16&2.173&20000&1000&--0.017&0.025\\
--3.51&1.453&18887&4571&0.079&0.030\\
--2.98&2.026&18315&2734&--0.046&0.029\\
--2.57&1.974&20000&1000&--0.014&0.025\\
--2.22&1.664&20000&1000&--0.027&0.025\\
--1.87&1.470&19261&2173&--0.075&0.027\\
--1.52&1.398&18994&2404&--0.072&0.027\\
--1.17&1.307&20000&1000&0.047&0.021\\
--0.88&1.615&20000&1000&--0.045&0.024\\
--0.64&1.321&20000&1000&--0.010&0.026\\
--0.41&1.315&20000&1000&--0.023&0.025\\
--0.18&1.448&18876&2505&--0.088&0.031\\
0.00&1.348&19196&1837&--0.115&0.028\\
0.29&1.334&20000&1000&0.003&0.021\\
0.47&1.991&20000&1000&--0.036&0.025\\
0.64&1.425&18638&2786&--0.054&0.032\\
0.88&1.288&20000&1000&--0.002&0.021\\
1.11&1.168&20000&1000&--0.050&0.025\\
1.35&1.399&17074&3735&--0.087&0.030\\
1.58&1.176&17165&3930&--0.091&0.031\\
1.87&1.144&14961&2016&--0.066&0.046\\
2.22&1.118&16172&2802&--0.077&0.040\\
2.63&1.213&16288&2747&--0.030&0.040\\
3.10&0.913&18197&2738&--0.088&0.035\\
3.57&0.950&13814&1870&--0.012&0.044\\
4.04&1.048&20000&1000&--1.249&0.022\\
4.50&1.201&20000&1000&0.001&0.021\\
5.03&1.152&20000&1000&--0.017&0.026\\
5.62&1.611&16303&2973&--0.005&0.041\\
6.32&1.464&17124&3720&--0.076&0.032\\
7.25&1.704&18521&2811&--0.069&0.034\\
8.54&2.212&1780&76&0.604&0.020\\
\hline
\end{tabular}
\caption{SSP-equivalent ages and metallicities fitted to MCG-02-02-086.}
\label{table:SSP_MCG}
\end{table*}

\begin{table*}
\centering
\begin{tabular}{r c r@{$\pm$}l r@{$\pm$}l}
\hline Radius  & \multicolumn{1}{c}{$\chi^{2}$} & \multicolumn{2}{c}{Age} & \multicolumn{2}{c}{[Fe/H]} \\
(arcsec) & \multicolumn{1}{c}{} & \multicolumn{2}{c}{(Myr)} & \multicolumn{2}{c}{} \\
\hline --4.09&2.992&14650&148&0.630&0.022\\
--2.28&1.923&16531&2470&0.041&0.032\\
--1.70&1.881&18699&2801&--0.036&0.033\\
--1.23&1.691&16146&2447&0.025&0.032\\
--0.94&1.509&16245&2687&0.127&0.034\\
--0.41&1.056&12700&2248&0.124&0.037\\
0.00 &1.586&16064&2896&0.028&0.031\\
0.41&1.002&13529&1795&0.013&0.026\\
0.94&1.035&6790&262&0.186&0.027\\
1.23&1.200&11453&1272&--0.003&0.031\\
1.46&1.381&6233&226&0.140&0.026\\
1.70&1.540&11549&1360&0.038&0.030\\
1.87&1.370&6554&202&0.167&0.025\\
2.05&1.137&15570&1960&0.024&0.026\\
2.28&1.120&7182&651&0.130&0.030\\
2.52&0.971&6437&150&0.221&0.027\\
2.75&0.959&16001&2306&0.057&0.042\\
2.98&1.086&6469&151&0.246&0.035\\
3.39&1.249&14610&1839&--0.037&0.029\\
3.74&2.152&9929&1694&0.043&0.033\\
4.09&2.281&20000&1000&--0.008&0.038\\
4.45&2.713&15995&2588&0.090&0.022\\
4.80&3.048&11916&907&0.548&0.034\\
5.15&0.801&20000&1000&--0.002&0.029\\
5.50&0.994&7284&1408&0.040&0.045\\
5.85&0.906&7385&1560&0.055&0.050\\
\hline
\end{tabular}
\caption{SSP-equivalent ages and metallicities fitted to PGC023233.}
\label{table:SSP_PGC}
\end{table*}

\begin{table*}
\centering
\begin{tabular}{r c c r@{$\pm$}l r@{$\pm$}l c}
\hline Radius  & \multicolumn{1}{c}{$\chi^{2}$} & Number of components & \multicolumn{2}{c}{Age} & \multicolumn{2}{c}{[Fe/H]} & Percentage Light-weighted. \\
(arcsec) & \multicolumn{1}{c}{} & & \multicolumn{2}{c}{(Myr)} & \multicolumn{2}{c}{} & \\
\hline --3.22&1.607&2&7159& 3505&0.232&0.171&50.0\\
  & & &12911&17844&--2.037&0.257&50.0\\
--1.70&1.519&2&16&14&--1.234&1.499&23.0\\
& & &7165&2367&0.162&0.059&77.0\\
--0.94&1.040&2&10&100&--2.100&0.915&21.5\\
& & &7139&1835&0.213&0.052&78.5\\
0.29&1.106&2&10&100&--2.301&1.000&24.4\\
& & & 7170&1913&0.164&0.056&75.6\\
0.64&0.826&2&13&8&--2.034&0.983&26.1\\
& & &7225&4191&0.198&0.181&74.1\\
1.11&1.085& 3&16&24&--0.995&1.851&7.9\\
& & &7178&2774&0.165&0.088&70.6\\
& & &20000&2000&--2.301&1.000&21.4\\
1.76&1.325&1&14546&3415&--0.423&0.076&100.0\\
3.22&2.202&1&9892&2139&--0.463&0.076&100.0\\
\hline
\end{tabular}
\caption{CSP ages and metallicities fitted to 2MASXJ17122774-2322108.}
\label{table:CSP_2MAS}
\end{table*}

\begin{table*}
\begin{tiny}
\centering
\begin{tabular}{r c c r@{$\pm$}l r@{$\pm$}l c}
\hline Radius  & \multicolumn{1}{c}{$\chi^{2}$} & Number of components & \multicolumn{2}{c}{Age} & \multicolumn{2}{c}{[Fe/H]} & Percentage Light-weighted. \\
(arcsec) & \multicolumn{1}{c}{} & & \multicolumn{2}{c}{(Myr)} & \multicolumn{2}{c}{} & \\
\hline --9.24 & 1.753 &2 & 1665& 2918&--2.291&1.238&11.0\\
& & & 18375&6844&0.118&0.047&89.0\\
--7.25 & 1.245&2&86&66&--2.301&1.000&16.4\\
& & & 20000&1000&0.079&0.037&83.6\\
--6.26 & 1.280&2&10&10&--1.933&1.817&15.3\\
& & & 20000&1000&--0.009&0.041&84.7\\
--5.62 & 1.158&2&46&50&--2.301&1.000&13.6\\
& & &19075&1056&0.117&0.045&86.4\\
--5.15 & 1.213&2&4192&2515&--1.642&1.084&31.9\\
& & &20000&1000&0.640&0.043&68.1\\
--4.80 & 1.172&2&10&10&--1.826&0.993&16.4\\
& & &20000&1000&0.092&0.035&83.6\\
--4.45 & 0.892&2&12&14&--1.707&1.537&17.0\\
& & &20000&1000&0.013&0.110&83.0\\
--3.92 & 1.010&2&25&21&--2.301&1.000&17.5\\
& & &17850&2950&0.216&0.045&82.5\\
--3.69 & 0.892&2&12&9&--1.893&0.895&13.5\\
& & &20000&1000&0.027&0.088&86.5\\
--3.45 & 0.772&2&23&16&--2.301&1.000&19.0\\
& & &17967&2672&0.295&0.047&81.0\\
--3.22 & 0.836&2&10&10&--1.875&1.511&16.6\\
& & &18337&5915&0.162&0.050&83.4\\
--3.00 & 0.789&2&271&364&--2.301&1.000&9.0\\
& & &17920&7133&0.139&0.477&91.0\\
--2.75 & 0.656&2&23&13&--2.301&1.000&18.2\\
& & &20000&1000&0.024&0.066&81.8\\
--2.52 & 0.752&2&10&10&--1.926&1.455&18.5\\
& & &18224&2983&0.291&0.050&81.5\\
--2.28 & 0.639&2&10&10&--1.912&1.048&18.5\\
& & &20000&1000&0.085&0.059&81.5\\
--2.05 & 0.616&2&25&18&--2.301&1.000&16.9\\
& & &18238&5483&0.161&0.040&83.1\\
--1.81 & 0.604&2&10&10&--1.938&1.471&18.8\\
& & &18604&6650&0.153&0.047&81.2\\
--1.58 & 0.615&2&14&22&--1.619&2.739&12.9\\
& & &20000&1000&0.039&0.100&87.1\\
--1.40 & 0.595&2&11&9&--1.804&2.716&14.2\\
& & &19201&7353&0.150&0.064&85.8\\
--1.29 & 0.813&2&294&176&--2.301&1.000&12.6\\
& & &18376&7278&0.164&0.064&87.4\\
--1.17 & 0.712&2&26&17&--2.301&1.000&17.4\\
& & &18060&3925&0.214&0.043&82.6\\
--1.05 & 0.586&2&1039&879&--2.301&1.000&8.7\\
& & &17848&4294&0.106&0.071&91.3\\
--0.94 & 0.763&2&10&10&--1.918&1.270&21.0\\
& & &18362&4096&1.270&0.231&79.0\\
--0.82 & 0.896&2&10&10&--0.408&0.338&19.5\\
& & &17839&2230&0.279&0.054&80.5\\
--0.70 & 0.692&2&10&10&--1.923&1.777&15.0\\
& & &18323&4399&0.200&0.044&85.0\\
--0.58 & 0.567&2&26&23&--2.301&1.000&13.6\\
& & &20000&1000&0.075&0.068&82.4\\
--0.41 & 0.643&2&30&29&--2.301&1.000&12.2\\
& & &20000&1000&0.081&0.033&87.8\\
--0.18 & 0.697&2&97&72&--1.382&1.153&16.3\\
& & &18279&9589&0.150&0.110&83.7\\
0.00 & 0.542&2&27&17&--2.301&1.000&13.9\\
& & &20000&1000&0.053&0.095&86.1\\
0.29 & 0.495&2&28&19&--2.301&1.000&15.2\\
& & &20000&1000&--0.005&0.151&84.8\\
0.53 & 0.525&2&27&15&--2.301&1.000&19.3\\
& & &17945&3333&0.223&0.045&80.7\\
0.76 & 0.504&2&10&10&--1.936&0.695&16.5\\
& & &20000&1000&--0.001&0.118&83.5\\
0.99 & 0.683&2&25&17&--2.301&1.000&17.0\\
& & &18954&5435&1.149&0.040&83.0\\
1.23 & 0.601 & 3&45&43&--2.301&1.000&12.9\\
& & &9517&8060&--0.259&1.047&20.9\\
& & &18895&4275&0.577&0.079&66.1\\
1.46 & 0.709&2&27&25&--2.301&1.000&11.9\\
& & &20000&1000&0.036&0.033&88.1\\
1.70 & 0.780&2&26&20&--2.301&1.000&15.3\\
& & &20000&1000&0.067&0.034&84.7\\
1.93 & 0.834&2&946&2782&--1.548&3.877&9.6\\
& & &18656&4024&0.205&0.072&90.4\\
2.16 & 0.891&2&13&10&--1.677&1.528&14.7\\
& & &20000&8730&0.051&0.056&85.3\\
2.40 & 0.912&2&10&10&0.699&1.000&23.4\\
& & &19908&3166&0.214&0.042&76.6\\
2.63 & 0.853&2&27&17&--2.301&1.000&14.7\\
& & &20000&1000&0.050&0.084&83.3\\
2.87 & 0.857&2&10&10&--2.301&1.000&30.1\\
& & &18783&4473&0.246&0.071&69.9\\
3.10 & 0.967&2&434&733&--2.301&1.000&13.3\\
& & &18102&4852&0.185&0.039&86.7\\
3.39 & 0.934&2&48&58&--2.301&1.000&12.7\\
& & &20000&1000&0.075&0.062&87.3\\
3.74 & 0.913&2&24&25&--2.301&1.000&13.5\\
& & &19353&1163&0.111&0.101&86.5\\
4.74 & 1.032&2&39&34&--2.301&1.000&12.2\\
& & &20000&1000&0.105&0.032&87.8\\
5.62 & 1.037&2&10&10&--2.085&1.117&22.1\\
& & &20000&1000&0.113&0.061&78.9\\
7.20 & 1.113&2&179&230&--2.301&1.000&14.2\\
& & &19464&1287&0.097&0.055&85.8\\
\hline
\end{tabular}
\caption{CSP ages and metallicities fitted to ESO541-013.}
\label{table:CSP_ESO}
\end{tiny}
\end{table*}

\begin{table*}
\begin{scriptsize}
\centering
\begin{tabular}{r c c r@{$\pm$}l r@{$\pm$}l c}
\hline Radius  & \multicolumn{1}{c}{$\chi^{2}$} & Number of components & \multicolumn{2}{c}{Age} & \multicolumn{2}{c}{[Fe/H]} & Percentage Light-weighted. \\
(arcsec) & \multicolumn{1}{c}{} & & \multicolumn{2}{c}{(Myr)} & \multicolumn{2}{c}{} & \\
\hline --8.42 & 2.430&2&433&1169&--0.426&2.046&10.2\\
 & & &19998&2412&--0.139& 0.102&89.8\\
--6.20& 2.343&2&422&1636&--0.143&3.123&6.4\\
& & &20000&1000&--0.011& 0.063&93.6\\
--4.97& 2.197&1&20000&1000&--0.129&0.024&100.0\\
--4.16& 2.132&2&10&100&0.699&0.100&14.5\\
& & &20000&1000&0.055&0.032&85.5\\
--3.51& 1.477&2&984&1172&--2.301&0.100&5.6\\
& & &18588&5136&0.142&0.036&94.4\\
--2.98& 1.927&2&421&760&--0.356&1.533&10.9\\
& & &20000&1000&0.043&0.055&89.1\\
--2.57& 1.848&2&10&100&0.699&0.100&19.5\\
& & &18715&6040&0.138&0.040&80.5\\
--2.22& 1.633&2&364&963&--2.301&0.100&6.7\\
& & &20000&1000&0.010&0.026&93.7\\
--1.87& 1.483&2&689&4658&--0.469&0.874&3.8\\
& & &20000&1000&--0.031&0.076&96.2\\
--1.52& 1.398&2&1018&838&--2.301&0.100&7.6\\
& & &20000&1000&0.001&0.042&92.4\\
--1.17& 1.320&2&314&1299&--2.298&0.926&5.3\\
& & &18967&7059&0.109&0.041&94.7\\
--0.88& 1.602&1&20000&1000&--0.034&0.046&100.0\\
--0.64& 1.311&1&20000&1000&--0.004&0.035&100.0\\
--0.41& 1.316&1&20000&1000&--0.031&0.031&100.0\\
--0.18& 1.433&2&693&5963&--0.701&0.226&3.4\\
& & &20000&1000&--0.047&0.087&96.6\\
0.00 & 1.350&2&544&6806&--1.378&0.720&1.6\\
& & &19740&3117&--0.088&0.081&98.4\\
0.29& 1.362&2&603&2830&--2.301&0.100&4.6\\
& & &20000&1000&0.037&0.041&95.4\\
0.47& 2.000&2&30&83&0.699&0.100&4.4\\
& & &20000&1000&--0.006&0.030&95.6\\
0.64& 1.424&2&857&2060&--2.301&0.100&5.4\\
& & &20000&1000&--0.001&0.030&94.6\\
0.88& 1.294&2&447&1864&--1.198&0.883&4.4\\
& & &20000&1000&0.039&0.050&95.6\\
1.11& 1.197&2&693&1951&--1.667&0.578&4.7\\
& & &20000&1000&--0.008&0.062&95.3\\
1.35& 1.322&2&1056&605&--2.301&0.100&11.3\\
& & &20000&1000&0.036&0.058&88.7\\
1.58& 1.186&2&969&576&--2.301&0.100&9.3\\
& & &20000&1000&0.000&0.028&90.7\\
1.87& 1.058&2&45&40&0.571&0.671&13.0\\
& & &20000&1000&0.037&0.034&87.0\\
2.22& 1.059&2&10&100&0.699&0.100&18.0\\
& & &20000&1000&0.088&0.035&82.0\\
2.63& 1.146&2&97&236&--1.707&0.300&7.3\\
& & &20000&1000&0.027&0.030&92.7\\
3.10& 0.899&2&1101&1416&--2.301&0.100&6.1\\
& & &20000&1000&--0.014&0.078&93.9\\
3.57& 0.895&2&83&75&--2.301&0.100&11.1\\
& & &17901&4854&0.095&0.037&88.9\\
4.04& 1.064&2&170&530&--0.895&0.363&6.9\\
& & &20000&1000&0.079&0.034&93.1\\
4.50& 1.197&2&547&1064&--1.603&0.239&6.7\\
& & &18771&6648&0.912&0.079&93.3\\
5.03& 1.166&2&508&852&--1.691&0.347&6.2\\
& & &20000&1000&0.014&0.063&93.8\\
5.62& 1.546&2&87&78&--2.301&0.100&15.3\\
& & &19034&8377&0.123&0.050&84.7\\
6.32& 1.425&2&953&667&--2.301&0.100&7.4\\
& & &20000&1000&0.006&0.027&92.6\\
7.25& 1.644&2&409&675&--0.338&1.463&12.7\\
& & &20000&1000&0.018&0.058&87.3\\
8.54& 2.105&2&54&26&0.632&0.295&24.8\\
& & &18777&3812&0.214&0.048&75.2\\
\hline
\end{tabular}
\caption{CSP ages and metallicities fitted to MCG-02-02-086.}
\label{table:CSP_MCG}
\end{scriptsize}
\end{table*}

\begin{table*}
\centering
\begin{tabular}{r c c r@{$\pm$}l r@{$\pm$}l c}
\hline Radius  & \multicolumn{1}{c}{$\chi^{2}$} & Number of components & \multicolumn{2}{c}{Age} & \multicolumn{2}{c}{[Fe/H]} & Percentage Light-weighted. \\
(arcsec) & \multicolumn{1}{c}{} & & \multicolumn{2}{c}{(Myr)} & \multicolumn{2}{c}{} & \\
\hline --4.09& 2.887&1&15000&2013&0.565&0.056&100\\
--2.28& 1.969&3&1996&3534&--2.225&1.795&5.1\\
& & &15657&11466&--0.261&0.263&48.2\\
& & &18985&4827&0.699&0.100&46.7\\
--1.70&1.863&1&18906&3604&--0.016&0.046&100.0\\
--1.23&1.716&3&15&38&--2.293&2.603&2.4\\
& & &5553&597&0.699&0.100&31.5\\
& & &15237&5014&--0.344&0.117&66.1\\
--0.94&1.524&2&3490&3572&--2.138&0.932&10.8\\
& & &17728&1798&0.225&0.057&89.2\\
--0.41&1.035&2&13289&1994&0.604&0.186&30.6\\
& & &17447&7607&--0.010&0.232&69.4\\
0.00 &1.681&3&1457&1173&--2.301&0.897&4.9\\
& & &5566&907&0.616&0.057&36.8\\
& & &20000&1000&--0.456&0.194&58.3\\
0.41&0.994&1&16445&6298&0.160&0.126&100\\
0.94&1.032&2&15409&764&0.699&0.100&53.5\\
& & &18724&6200&--0.304&0.138&46.5\\
1.23&1.164&1&17845&1652&0.222&0.059&100\\
1.46&1.324&2&15485&3385&0.435&0.073&70.9\\
& & &20000&2851&--1.221&0.540&29.1\\
1.70&1.449&3&1125&2532&--1.261&2.593&3.1\\
& & &10937&10483&0.308&0.251&44.5\\
& & &14581&15333&--0.410&0.475&52.4\\
1.87&1.340&2&2000&1000&--2.301&0.100&7.9\\
& & &16699&4022&0.142&0.034&92.1\\
2.05&1.126&2&1043&1091&--1.202&2.232&5.4\\
& & &19782&3813&--0.112&0.192&94.6\\
2.28&1.068&3&164&119&0.699&0.100&3.9\\
& & &17753&3748&0.664&0.071&39.0\\
& & &20000&1000&--0.178&0.163&57.0\\
2.52&0.913&3&287&397&--2.301&0.100&6.0\\
& & &4559&5052&0.697&0.625&1.0\\
& & &16689&6653&0.133&0.069&93.0\\
2.75&0.964&3&13&22&0.504&0.802&8.8\\
& & &15355&3355&0.660&0.451&25.0\\
& & &17119&7186&0.056&0.351&65.2\\
2.98&1.058&2&6962&8416&--2.240&0.568&14.8\\
& & &14028&2916&0.315&0.067&85.2\\
3.39&1.240&2&10&100&0.588&0.616&6.9\\
& & &17568&4033&0.005&0.054&93.1\\
3.74&2.110&3&139&2036&--2.291&1.497&1.0\\
& & &9180&6168&--2.187&0.371&26.8\\
& & &16273&5440&0.446&0.067&72.2\\
4.09&2.244&3&203&6561&--1.161&2.579&0.5\\
& & &11888&13042&--0.818&0.428&40.5\\
& & &18810&4744&0.632&0.0060&59.0\\
4.45&2.800&2&508&2003&--0.464&2.551&6.5\\
& & &16109&3997&--0.145&0.077&93.5\\
4.80&3.048&1&11916&907&0.548&0.034&100.0\\
5.15&0.801&1&20000&1000&--0.002&0.029&100.0\\
5.50&0.994&1&7284&1408&0.040&0.045&100.0\\
5.85&0.906&1&7385&1560&0.055&0.050&100.0\\
\hline
\end{tabular}
\caption{CSP ages and metallicities fitted to PGC023233.}
\label{table:CSP_PGC}
\end{table*}

\section{Conclusion}
\label{summary}
We have obtained, analysed and interpreted detailed, high spatial and spectral resolution, long-slit observations of four CCGs (Abell 0085, 0133, 0644 and Ophiuchus) recently obtained on SALT, to probe the spatially-resolved kinematics and stellar populations of the stars. We use the Pegase.HR model with the ELODIE v3.1 stellar library to determine the SFHs of the galaxies using the full spectrum fitting. We perform SSP as well as CSP fits to acount for more complex SFHs. Monte-Carlo simulations and $\chi^{2}$-maps are used to check the reliability of the solutions. We emphasize that we only interpret the presence of young stellar components which is much more reliable than interpreting the absolute ages, metallicities and $\alpha$-enhancements.

Hudson et al.\ (2010) use the largest complete sample of 64 galaxy clusters (HIghest X-ray FLUx Galaxy Cluster Sample) with available high-quality X-ray data from Chandra, and apply 16 cool-core diagnostics to them. To segregate cool core and non-cool-core clusters, they find that central cooling time  is the best parameter for low redshift clusters with high quality data. They further show that the discrepancy in classical and spectroscopic mass deposition rates can not be explained with a recent formation of the cool cores, demonstrating the need for a heating mechanism to explain the cooling flow problem.

Only a few CCGs have been reported in the literature with ongoing or recent bursts of star formation, and the majority -- if not all of them -- are hosted in 'cooling flow' clusters. Rafferty et al.\ (2008), Cavagnolo et al.\ (2008, 2009) and Hoffer et al.\ (2012) show that central galaxies in clusters with low entropy ($K_{0}$) -- thus with large cooling rates -- are the only CCGs to exhibit signs of vigorous star formation. Here, we detect very small, but observable young stellar components in two of our four clusters which have no mass deposition rates (or very little so that it is below the X-ray detection limit). This suggests another possible mechanism for the recent star formation such as galactic cannibalism that appears in cluster environments due to dynamical friction. We note that the spatially resolved recession velocity of these two galaxies (2MASXJ17122774-2322108 and ESO541-013) show measurable rotation (see Figure \ref{fig:velocity}), although the sample is too small to infer a possible difference between the kinematics of these galaxies and the galaxy with a younger component hosted by a cooling flow cluster. Abell 0133 (host of ESO541-013) showed extended H-alpha filaments in McDonald et al. (2010), whereas the other three clusters examined here showed nuclear H-alpha emission. Therefore no obvious correlation between the H-alpha gas and stellar kinematics exists. Mergers that are not dissipation-less can supply fresh cold gas that can trigger an episode of star formation (Mihos $\&$ Hernquist 1996). A large library of smooth particle hydrodynamic simulations of galaxy mergers (Di Matteo et al.\ 2007, 2008) show that the average enhancement of the star formation rate in a random galaxy collision is only a factor of a few (3-4 being the median factor) and only lasts 200-400 Myr. This process is more efficient in dense clusters but the star formation rate increase remains in general below a factor of 10 or less (Martig $\&$ Bournaud 2008).

Since we have carefully measured the possible presence of emission lines (with particular attention to the Balmer lines), and couldn't find any detections above three sigma of the noise of the spectra in any of the four objects - we conclude that our slit placement did not intercept the H$\alpha$ filaments. Thus in our data, the relatively young stars do not necessarily co-exist with the H$\alpha$ emitting gas.

This study, focussing on the detailed analysis of four objects, is complementary to our previous studies investigating the evolution and environmental influence in a larger sample of central cluster galaxies (Loubser et al.\ 2008, Loubser $\&$ Sanchez-Blazquez 2012). Throughout the investigation and analysis of the larger sample, a picture emerged indicating that the kinematics and stellar population properties of the central galaxies are not very dependent on the galaxy mass (with the exception of the metallicity gradients which are dependent on the galaxy velocity dispersions). Similarly, various cluster properties also seem to have very limited influence on the central galaxy stellar population properties, such as cluster temperature or density. On the other hand, the offset between the X-ray peak of the cluster and the galaxy and the cooling flows in the clusters do influence the probability that the central galaxy will form stars. In Loubser et al.\ (2008), we found that one of the six younger galaxies were hosted by a non-cooling flow cluster whereas the other five younger galaxies were hosted by a cooling-flow cluster. This previous study analysed only the SSP-equivalent ages, but reinforces the findings of the current study. All of these observations point to very diverse star formation histories of these central galaxies, and excludes the naive picture of very passive, homogeneous evolution.  

\section*{Acknowledgments}
SIL is financially supported by the South African National Research Foundation. SIL gratefully acknowledges the constructive comments from the anonymous reviewer as well as the help of Alet de Witt (Hartebeesthoek Radio Observatory) with the radio data interpretation. All of the observations reported in this paper were obtained with the Southern African Large Telescope (SALT) under program 2011-2-RSA$\_$OTH-003 and 2012-1-RSA$\_$OTH-003 (PI: Ilani Loubser).

\appendix

\clearpage

\section{Stellar kinematics}
\label{Kin_tables}

The spatially resolved kinematic data for all 4 CCGs are available as online-data.

\begin{figure*}
   \centering
   \includegraphics[scale=0.5, angle=270]{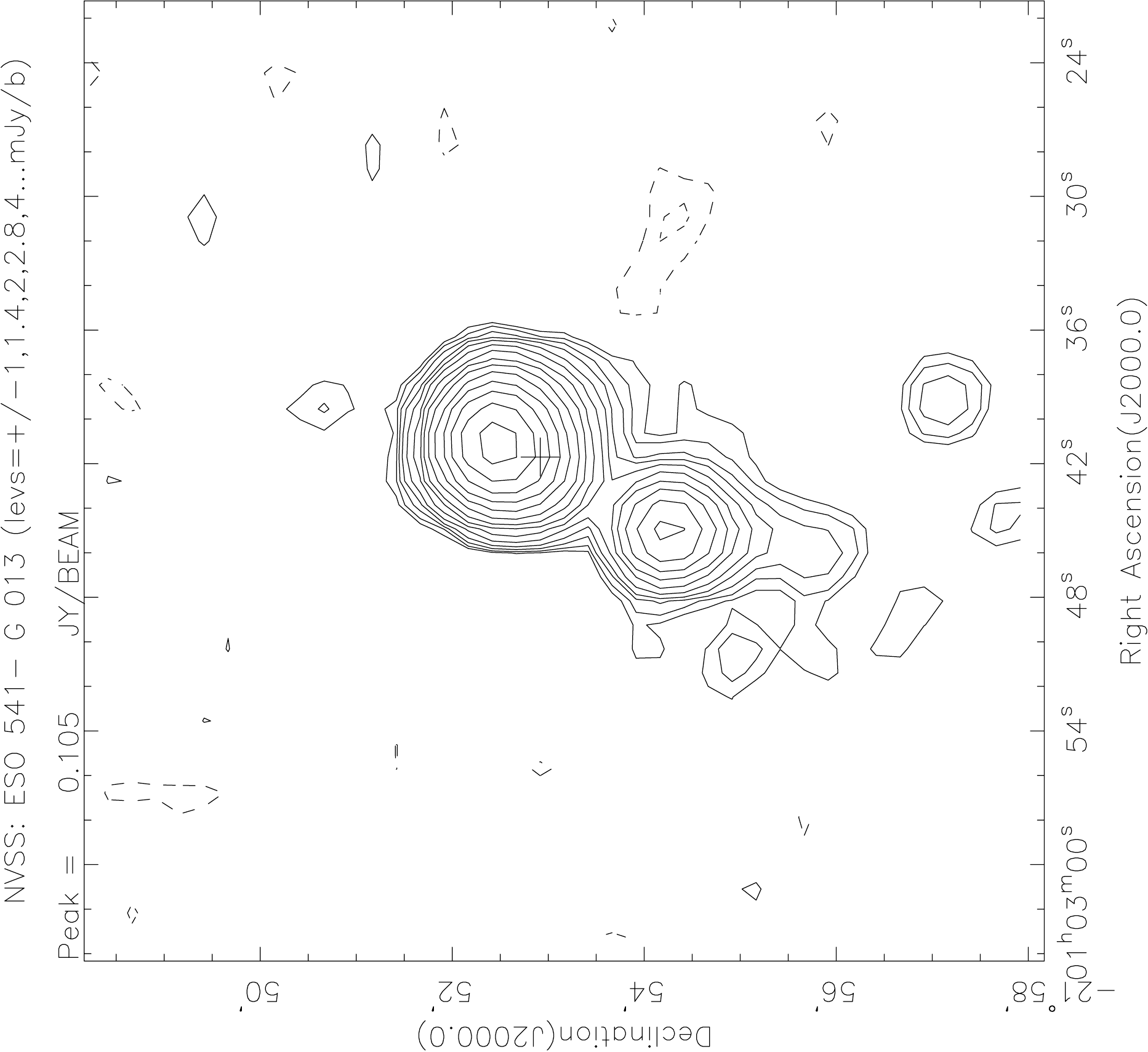}
   \caption{ESO541-013 NRAO VLA Sky Survey (NVSS) 10 $\times$ 10 arcmin image.}
   \label{fig:ESO_NVSS}
\end{figure*}

\begin{figure*}
   \centering
   \includegraphics[scale=0.5, angle=270]{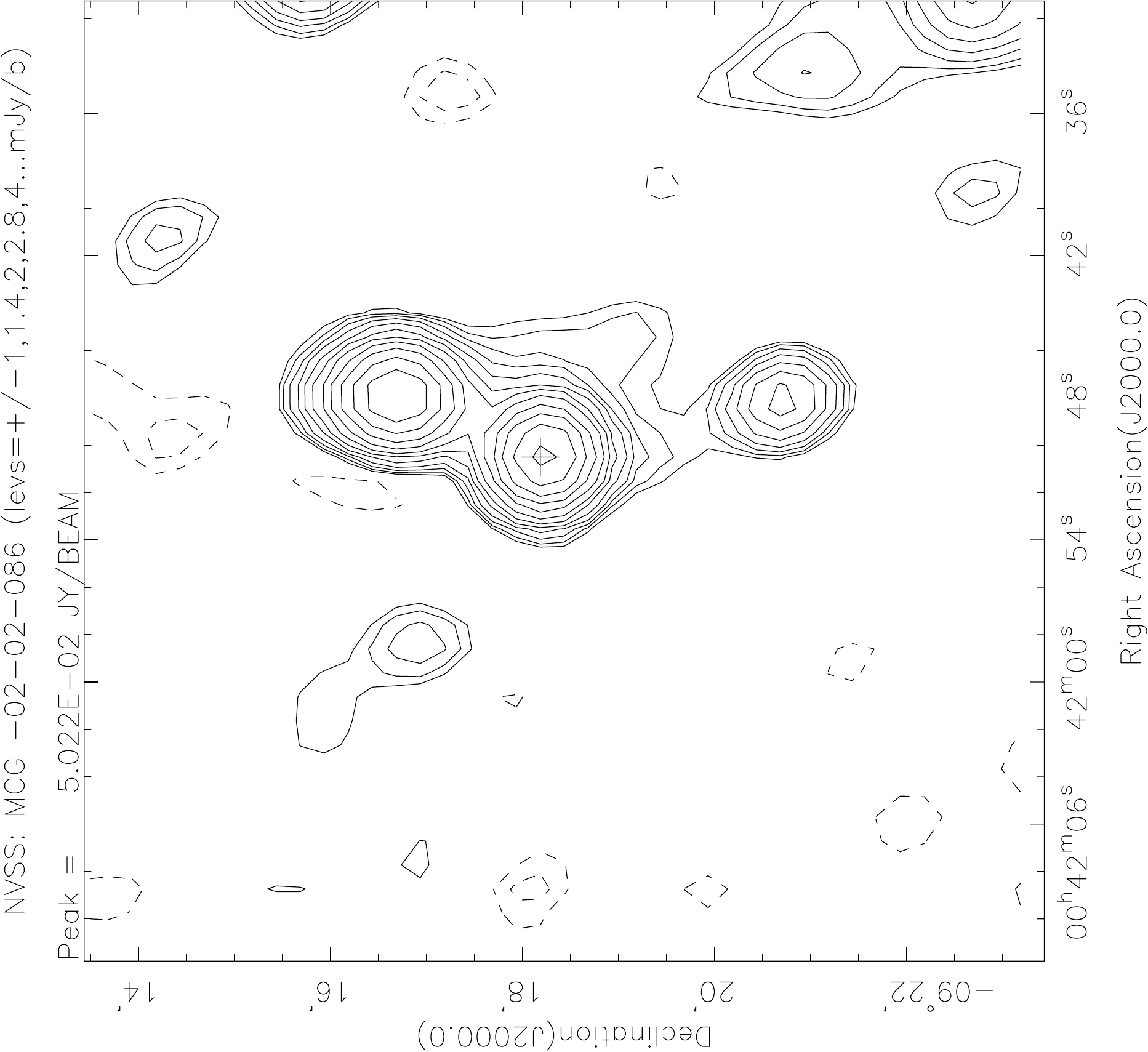}
   \caption{MCG-02-02-086 NVSS 10 $\times$ 10 arcmin image.}
   \label{fig:MCG_NVSS}
\end{figure*}

\begin{figure*}
   \centering
   \includegraphics[scale=0.5, angle=270]{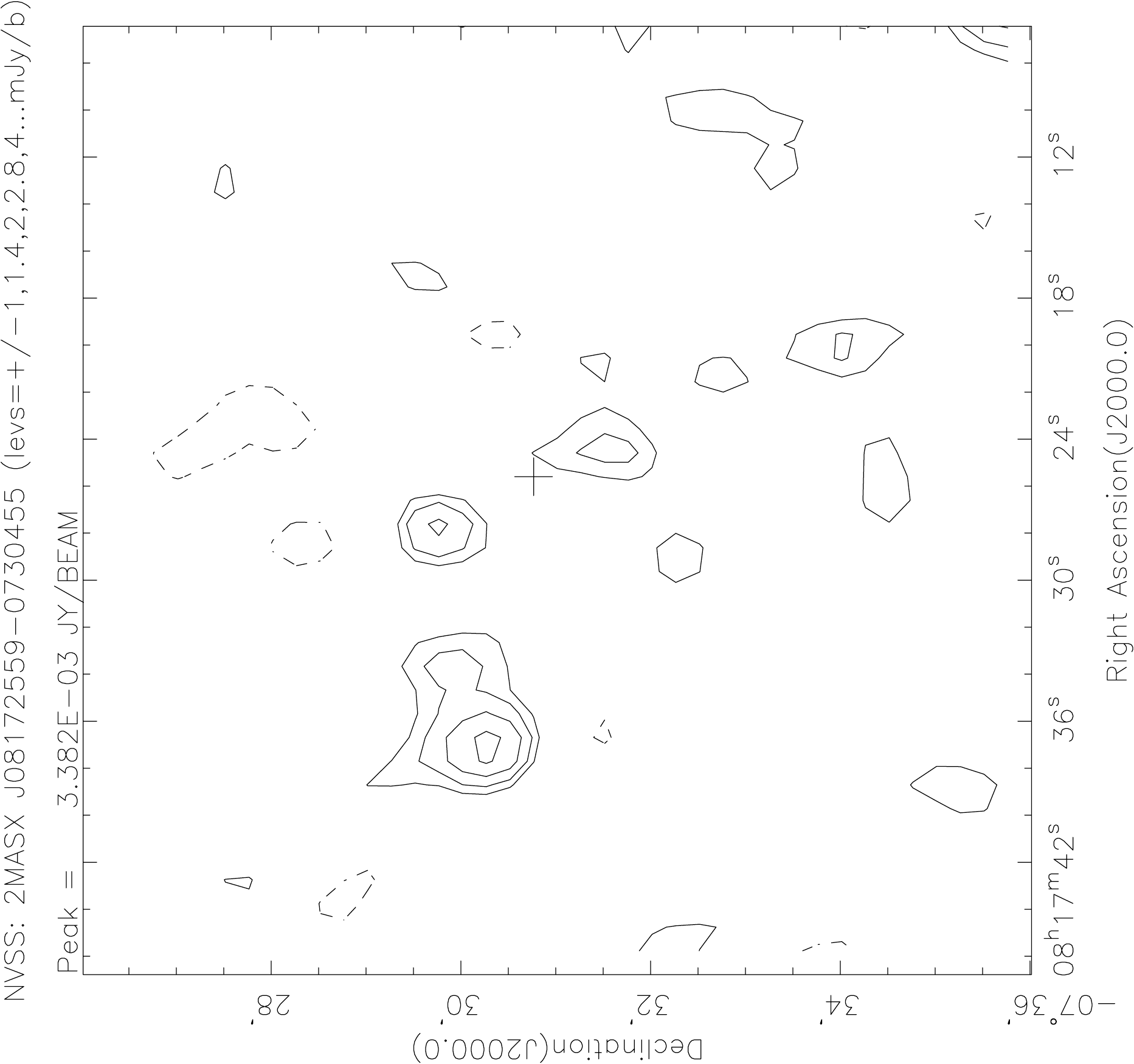}
   \caption{PGC023233 NVSS 10 $\times$ 10 arcmin image.}
   \label{fig:PGC_NVSS}
\end{figure*}

\begin{figure*}
   \centering
   \includegraphics[scale=0.5, angle=270]{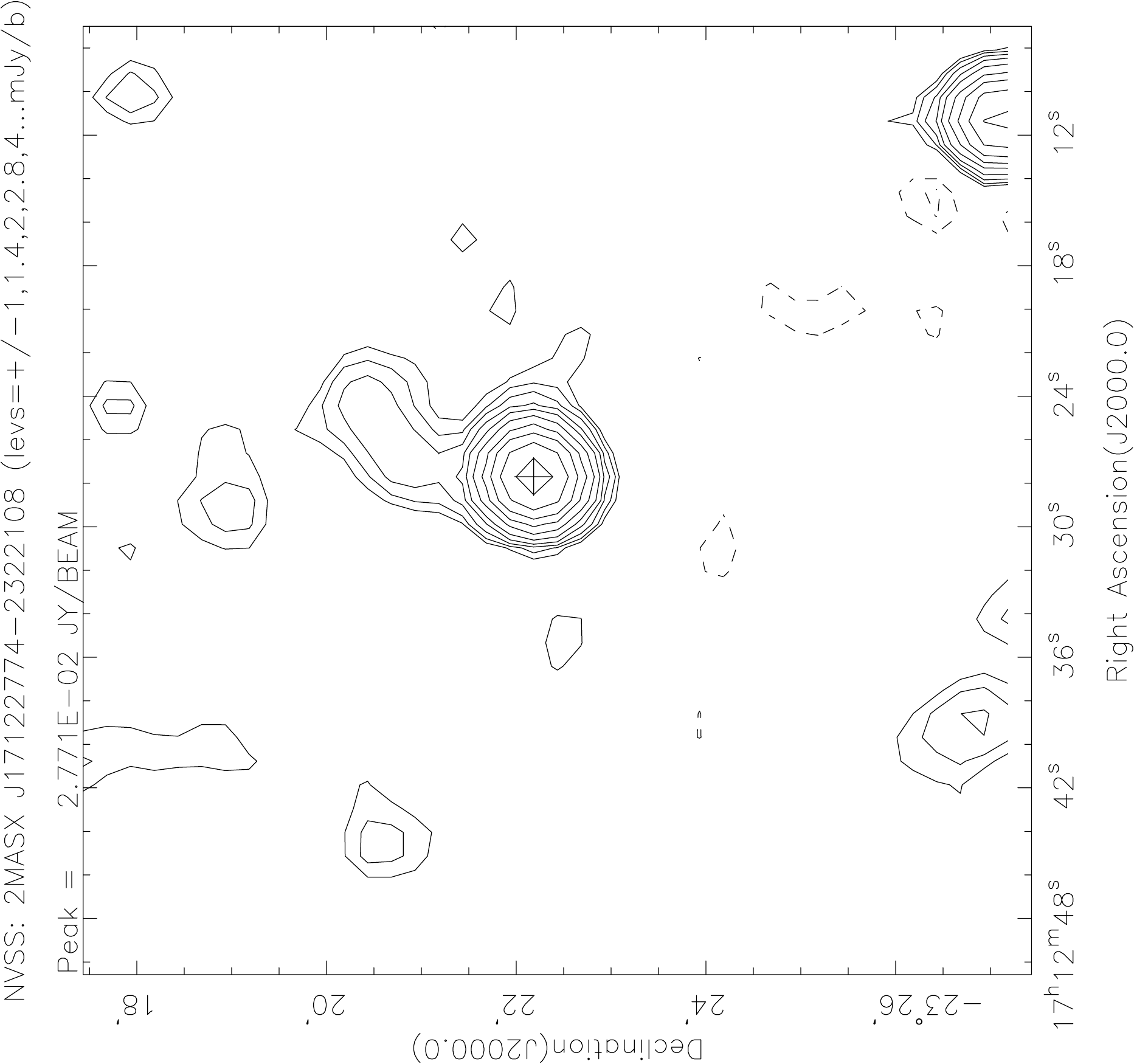}
   \caption{2MASXJ17122774-2322108 NVSS 10 $\times$ 10 arcmin image.}
   \label{fig:2MAS_NVSS}
\end{figure*}

\bsp

\label{lastpage}

\end{document}